\documentclass[sn-mathphys,Numbered]{sn-jnl}

\usepackage{graphicx}%
\usepackage{physics}
\usepackage{multirow}%
\usepackage{amsmath,amssymb,amsfonts}%
\usepackage{amsthm}%

\usepackage{mathrsfs}%
\usepackage[title]{appendix}%
\usepackage{xcolor}%
\usepackage{textcomp}%
\usepackage{manyfoot}%
\usepackage{booktabs}%
\usepackage{algorithm}%
\usepackage{algorithmicx}%
\usepackage{algpseudocode}%
\usepackage{listings}%

\usepackage{dcolumn}

\usepackage{bm}
\usepackage{hyperref}
\usepackage{bbold}
\usepackage{subfigure}
\usepackage{lineno}

\usepackage{etoolbox} 

\newcommand*\linenomathpatch[1]{%
  \cspreto{#1}{\linenomath}%
  \cspreto{#1*}{\linenomath}%
  \csappto{end#1}{\endlinenomath}%
  \csappto{end#1*}{\endlinenomath}%
}

\linenomathpatch{equation}
\linenomathpatch{gather}
\linenomathpatch{multline}
\linenomathpatch{align}
\linenomathpatch{alignat}
\linenomathpatch{flalign}

\begin{document}

\title[Article Title]{Digital Simulation of Convex Mixtures of Markovian and Non-Markovian Single Qubit Pauli Channels on NISQ Devices}


\author[1,2]{\fnm{I. J.} \sur{David}}
\email{ian.david@nithecs.ac.za}
\author*[1,2]{\fnm{I.} \sur{Sinayskiy}}
\email{sinayskiy@ukzn.ac.za}

\author[1,2,3]{\fnm{F.} \sur{Petruccione}}
\email{petruccione@sun.ac.za}

\affil[1]{School of Chemistry and Physics, University of KwaZulu-Natal, Durban 4001, South Africa}

\affil[2]{National Institute for Theoretical and Computational Sciences (NITheCS), Stellenbosch, South Africa.}

\affil[3]{School of Data Science and Computational Thinking and Department of Physics, Stellenbosch University, Stellenbosch 7604, South Africa.}

\abstract{Quantum algorithms for simulating quantum systems provide a clear and provable advantage over classical algorithms in fault-tolerant settings. There is also interest in quantum algorithms and their implementation in Noisy Intermediate Scale Quantum (NISQ) settings. In these settings, various noise sources and errors must be accounted for when executing any experiments. Recently, NISQ devices have been verified as versatile testbeds for simulating open quantum systems and have been used to simulate simple quantum channels. Our goal is to solve the more complicated problem of simulating convex mixtures of single qubit Pauli channels on NISQ devices. We consider two specific cases: mixtures of Markovian channels that result in a non-Markovian channel (M+M=nM) and mixtures of non-Markovian channels that result in a Markovian channel (nM+nM=M). For the first case, we consider mixtures of Markovian single qubit Pauli channels; for the second case, we consider mixtures of Non-Markovian single qubit depolarising channels, which is a special case of the single qubit Pauli channel. We show that efficient circuits, which account for the topology of currently available devices and current levels of decoherence, can be constructed by heuristic approaches that reduce the number of CNOT gates used in our circuit. We also present a strategy for regularising the process matrix so that the process tomography yields a completely positive and trace-preserving (CPTP) channel.}


\maketitle

\noindent \textbf{Abbreviations:}
\begin{itemize}
    \item NISQ - Noisy Intermediate Scale Quantum.
    \item M+M=nM - Convex sum of two Markovian channels leading to a non-Markovian channel.
    \item nM+nM=M - Convex sum of two non-Markovian channels leading to a Markovian channel.
    \item GKSL - Gorini-Kossakowski-Sudarshan-Lindblad.
    \item CPTP - Completely Positive and Trace Preserving.
    \item QPT - Quantum Process Tomography.
    \item IM - Intermediate Map.
    \item CP - Completely Postive.
    \item IBMQE - IBM Quantum Experience.
    \item CNOT - Controlled Not.

\end{itemize}

\textbf{Key Points:}

\begin{itemize}
    \item This work simulates the convex mixtures of single qubit Markovian and non-Markovian quantum channels on NISQ devices provided by the IMBQE.
    \item The circuits used to implement the channels take into account the topolgy of the quantum device used as well as the number of CNOT gates used.
    \item We present a strategy for regularising the process matrix to ensure the quantum process tomography yields a CPTP channel. Something that is not correctly implemented in Qiskit.
    \item A method is outlined for finding mixtures of non-Markovian depolarising channels that yield a Markovian depolarising channel. It is also shown that, one cannot convexly mix two Markovian depolarising channels that leads to a non-Markovian depolarising channel. 

\end{itemize}

\section{Introduction}

Simulating large, complex quantum systems with classical computers is a computationally challenging problem. The computational resources required to perform these simulations would scale exponentially with the number of quantum particles, and these simulations would quickly become classically intractable. For this reason, Quantum Simulation, the simulation of quantum systems with quantum computers, has been proposed \cite{feynman1982simulating,manin1980computable}. The computational resources required to use quantum computers would scale only polynomially with the number of quantum particles. Since this discovery, quantum simulation has become one of the main motivations for developing quantum computers \cite{lloyd1996universal}. \\
Researchers have developed many algorithms to simulate quantum systems using quantum computers \cite{childs2012hamiltonian,childs2018toward,childs2019faster,childs2021theory,berry2007efficient,berry2015hamiltonian,campbell2019random,papageorgiou2012efficiency,low2019hamiltonian,berry2020time}. However,  these algorithms are best suited for fault-tolerant settings. In these settings, quantum computers would provide a clear advantage over classical computers in the simulation of quantum systems. However, despite the advancements made in quantum hardware, much more work needs to be done to reach fault-tolerant settings. This, and the growing accessibility of Noisy Intermediate Scale Quantum (NISQ) computers through the cloud from platforms such as the IBM Quantum Experience (IBMQE), has inspired interest in the use of NISQ devices for simulating quantum systems \cite{lau2022nisq,lau2021quantum,daley2022practical,funcke2022towards}. 

 In these settings, the simulations are restricted by the size of the systems, various error rates and noise sources. Despite this, even currently available quantum computers provide versatile test beds for various theories in quantum physics \cite{gonzalez2022hardware,takeshita2020increasing,garcia2020ibm}. 
 
A majority of recent work has been devoted to the the simulation of closed quantum systems in the NISQ era \cite{lau2022nisq,lau2021quantum,daley2022practical,funcke2022towards}. However, there has been less work done in the simulation of open quantum systems in the NISQ era \cite{sun2021efficient,han2021experimental,garcia2020ibm}. Open quantum systems \cite{FP07, RH12}, are systems that are allowed to interact with their environment and to simulate them we need to be able to simulate their evolution on a quantum computer.

A master equation describes the dynamics of an open quantum system. The solution of this master equation is a dynamical map, also known as a quantum channel, which describes the evolution of an open quantum system. Under certain assumptions, such as the Born-Markov approximation, the Gorini-Kossakowski-Sudarshan-Lindblad (GKSL) form of the master equation can be derived \cite{gorini1976completely,lindblad1976generators}. The GKSL form of the master equation describes Markovian dynamics, where all memory effects are neglected. Non-Markovian dynamics differs from Markovian dynamics in that it allows information to flow back into the system from the environment and does not neglect memory effects \cite{caruso2014quantum}.   
While everyone agrees that the time-independent GKSL generator with non-negative damping constants describes the quantum Markov process, there are several approaches to defining the non-Markovianity in quantum physics \cite{breuer2009measure,rivas2010entanglement,li2018concepts, pollock2018operational}. This makes the study of non-Markovianity in quantum physics a highly non-trivial problem. \\
There has been much interest in the study of the non-Markovian dynamics of an open quantum system \cite{wolf2008assessing,li2018concepts,breuer2009measure,rivas2010entanglement,chruscinski2013non,wudarski2016markovian,chruscinski2015non,vacchini2012classical,hall2014canonical}. There are many different descriptions of non-Markovianity \cite{rivas2010entanglement,breuer2009measure,hall2014canonical} however, in this work, we make use of CP divisibility \cite{rivas2010entanglement} to characterise a channel as Markovian or non-Markovian. We say that a channel is Markovian if it is CP-divisible and non-Markovian if it is CP-indivisible. 

Recently, simulations of open quantum systems have been used to understand non-Markovian dynamics further \cite{uriri2020experimental,garcia2020ibm}. 
There is a need for more algorithms for simulating open quantum systems with NISQ devices and more experiments to test existing algorithms \cite{garcia2020ibm}. This would demonstrate that NISQ devices can be used for more practical problems, taking us further toward using quantum computers as practical computing devices. 

Recently, there has been much interest in studying the mixtures of quantum channels and how mixing channels with specific properties leads to a new channel with counterintuitive and surprising properties \cite{siudzinska2020quantum,siudzinska2022non,siudzinska2022phase}. In \cite{siudzinska2022phase}, the author studies mixtures of commutative, unital and Markovian quantum channels, and they show that Markovianity can be recovered by mixing a certain number of phase covariant channels. The work done in \cite{siudzinska2020quantum} calculates the number of negative decay rates that qudit channels can have while still being physically legitimate quantum channels. The author of \cite{siudzinska2022non} provides connections between the non-Markovianity degree of general phase damping qubit maps and their legitimate mixtures. Necessary and sufficient conditions for Pauli maps to satisfy divisibility criteria are formulated. In \cite{siudzinska2020quantum}, mixtures of Pauli channels are studied, but a prescription for how to design these mixtures of channels for simulations and experimental demonstrations is not provided. In this work, we will provide a prescription for designing mixtures of depolarising channels to mix two non-Markovian depolarising channels and yield a Markovian channel. We also show that it is impossible to mix two Markovian depolarising channels and obtain a non-Markovian channel. Of interest to us is the work done by \cite{uriri2020experimental}, where they study mixtures of Markovian channels that lead to Non-Markovian channels and vice versa. 

In this work, we use NISQ devices, provided by the IBMQE through the cloud, to simulate the mixtures of Markovian quantum channels that yield a Non-Markovian channel (M+M=nM) and the mixtures of Non-Markovian channels that yield a Markovian channel (nM+nM=M). We will also use the Python package Qiskit to interact with the NISQ devices \cite{Qiskit}. For the case of M+M=nM, we consider mixtures of Pauli channels because of their relevance to bit-flip and phase-flip error encountered on quantum devices, while for the case of nM+nM=M, we consider mixtures of depolarising channels because of the common occurrence of depolarising noise on quantum devices. 

Each mixture was then simulated on a NISQ device. We show that efficient circuits that account for the topology of currently available devices and current levels of decoherence can be constructed by carefully considering device properties when applying the unitary dilation, such as Stinespring dilation \cite{stinespring1955positive}. These NISQ device motivated circuit constructions provide an improvement in the quality of the results obtained in simulations.

The channel that resulted from the simulation was then reconstructed. Previous attempts at simulating convex mixtures employed Maximum Likelihood Estimation (MLE) \cite{uriri2020experimental} to reconstruct the channel. This has since been shown, experimentally, to be unideal. In this work, we reconstruct the channel by solving a convex optimisation problem with problem-specific constraints \cite{huang2020reconstruction}. With this approach, a CPTP channel is obtained. The simulation was then verified by characterising the reconstructed channel as either Markovian or Non-Markovian using the CP-divisibility criteria \cite{rivas2010entanglement}. We also show that the reconstructed channels have high fidelity to the theoretical channel and have Choi matrices with the required trace norm of one. This indicates that the channels simulated are physical.
The rest of this paper is outlined as follows. In section 2, we introduce some background information and theory related to quantum channels and their generators, and we also present the theoretical design of the quantum channels to be simulated for both cases M+M=nM and nM+nM=M. In section 3, we discuss the simulation of the channels on the NISQ device. We discuss the quantum circuits designed to simulate the quantum channels in section 3 A. In section 3 B, we outline the method for process tomography and the convex optimisation technique we use to reconstruct a CPTP map. Section 4 will describe the method to characterise the simulated channels as Markovian or non-Markovian using the CP-divisibility criteria. The results from the simulation will be presented in detail in section 5. Lastly, in section 6, we make some concluding remarks and discuss possible extensions of this work.

\section{Preliminaries}

The evolution of an isolated (closed) quantum system is described by the von-Neumann equation,
\begin{align}
\label{eqvonNeumann}
    \dot{\rho}_{S}(t)=-i[H_{S},\rho_{S}(t)]
\end{align}
where $H_{S}$ is the Hamiltonian of the system and $\rho_{S}(t)$ is the density matrix of the system at some time $t\geq 0$. Formally, the solution to equation (\ref{eqvonNeumann}) is a unitary transformation,
\begin{align}
    \rho_{S}(t)=U(t)\rho_{S}(0)U^{\dagger}(t),
\end{align}
where $U(t)=\exp(-itH_{S})$. If the system is not isolated and is allowed to interact with some environment then the system and environment together undergo unitary evolution as an isolated system. To describe the dynamics of the system we need to "trace out" the environmental degrees of freedom obtaining the evolution of the open quantum system. Without loss of generality one can assume that the total state $\rho_{\mathrm{tot}}(t)$, at $t=0$, of the system and environment is a product state i.e. 
\begin{align}
    \rho_{\mathrm{tot}}(0)=\rho_{E}(0)\otimes \rho_{S}(0).
\end{align}
where $\rho_{E}$ is the density matrix of the environment and $\rho_{S}$ is the density matrix of the system. The total unitary evolution $U_{\mathrm{tot}}(t)$ of  the system and environment is given by, $U_{\mathrm{tot}}(t)=\exp(-itH_{\mathrm{tot}})$
where $H_{tot}$ is the Hamiltonian of the system and environment and is,
\begin{align}
    H_{\mathrm{tot}}=H_{S}+H_{E}+H_{I}
\end{align}
where $H_{S}$ is the system Hamiltonian, $H_{E}$ is the Hamiltonian of the environment and $H_{I}$ is the Hamiltonian describing the interaction between the system and environment. The total system and environment undergo unitary evolution i.e. $\rho_{\mathrm{tot}}(t)=U_{\mathrm{tot}}(t)\rho_{\mathrm{tot}}(0)U_{\mathrm{tot}}^{\dagger}(t)$, to obtain the dynamics of just the system we must trace out the environment using the partial trace so that,
\begin{align}
\label{eqfullevol}
    \rho_{S}(t)=\mathrm{tr}_{E}\left(U_{\mathrm{tot}}(t)\rho_{\mathrm{tot}}(0)U_{\mathrm{tot}}^{\dagger}(t)\right)
\end{align}
Equation (\ref{eqfullevol}) allows us to describe the effective dynamics of the system using a dynamical map $\Lambda_{t}$ where,
\begin{align}
    \Lambda_{t}\rho_{S}(0)=\mathrm{tr}_{E}\left(U_{\mathrm{tot}}(t)\rho_{\mathrm{tot}}(0)U_{\mathrm{tot}}^{\dagger}(t)\right)
\end{align}

The dynamical map $\Lambda_{t}$ where $t \geq 0$ and  $\Lambda_{0}=\mathbb{1}$  are a family of single parameter completely positive and trace preserving (CPTP) maps. From this point on since we are only interested in the dynamics of the system we can drop the subscript and represent the state of the system at some time $t\geq 0$, with just $\rho(t)$ so that, if $\rho(0)$ is the initial state of the system then $\rho(t)=\Lambda_{t}\rho(0)$  \cite{FP07}.

A dynamical map is also referred to as a quantum channel. These shall be used interchangeably throughout this work. One can assume that the map $\Lambda_{t}$, in  most practical cases, satisfies the time-local master equation,
\begin{equation}
\label{time_local_ME}
\frac{d}{dt}\rho(t)=\mathcal{L}(t)\rho(t)   \Leftrightarrow  \frac{d}{dt}\Lambda_{t}=\mathcal{L}(t)\Lambda_{t},
\end{equation}
where $\mathcal{L}(t)$ is the time-local generator and has the known form,
\begin{align}
    \mathcal{L}(t)\rho &=-i\big[ \hat{H}(t),\rho\big]+\sum_{k}\gamma_{k}(t)\bigg(\hat{V}_{k}(t)\rho\hat{V}_{k}^{\dagger}(t)-\frac{1}{2}\{\hat{V}_{k}^{\dagger}(t)\hat{V}_{k}(t),\rho \} \bigg),
\end{align}
where $\hat{H}(t)$ is the time-dependent Hamiltonian, $\gamma_{k}$(t) and $\hat{V_{k}}(t)$ are the time-dependent decay rates and noise operators respectively. The form of the generator in (2) is very general and gives both Markovian and non-Markovian dynamics. It is widely accepted that the famous Gorini-Kossakowski-Sudarshan-Lindblad (GKSL) form of the generator corresponds to Markovian dynamics \cite{gorini1976completely,lindblad1976generators}. The GKSL form of the generator is,
    \begin{align}
            \mathcal{L}\rho &=-i\big[ \hat{H},\rho\big]+\sum_{k}\gamma_{k}\bigg(\hat{V}_{k}\rho\hat{V}_{k}^{\dagger}-\frac{1}{2}\{\hat{V}_{k}^{\dagger}\hat{V}_{k},\rho \} \bigg) .
    \end{align}
Since the GKSL form of the generator corresponds to Markovian dynamics, all memory effects are neglected. A crucial part of this work will involve classifying the simulated dynamical maps $\Lambda_{t}$ as either Markovian or non-Markovian. As stated above, we shall use the CP divisibility criteria to classify a dynamical map as Markovian or non-Markovian \cite{rivas2010entanglement}, which shall be outlined in more detail in the next section. Another approach in studying the Markovianity of a dynamical map is to model the total system and environment dynamics, this work does not take this approach, but one could look at  \cite{li2018concepts,pollock2018operational} for more information.
The aim of this work can be stated formally as follows. For two channels $\Lambda_{t}^{(1)}$ and $\Lambda_{t}^{(2)}$ we are interested in classifying the total channel:
\begin{equation}
    \Lambda_{t}^{(T)}=\eta\Lambda_{t}^{(1)} + (1-\eta)\Lambda_{t}^{(2)}
\end{equation}
where $\eta \in [0,1]$. We consider the following two cases:\\
(i) $\Lambda_{t}^{(1)}$, $\Lambda_{t}^{(2)}$ are Markovian and $\Lambda_{t}^{(T)}$ is non-Markovian, which we shall refer to as (M+M=nM).\\
(ii) $\Lambda_{t}^{(1)}$, $\Lambda_{t}^{(2)}$ are non-Markovian and $\Lambda_{t}^{(T)}$ is Markovian, which shall be referred to as (nM+nM=M).\\
These two cases above arise from the non-convex geometry of the set of Markovian and non-Markovian channels \cite{wolf2008assessing}. Past works that have studied this non-convex geometry have provided examples of mixtures of Markovian channels leading to a non-Markovian channel, and vice versa \cite{wudarski2016markovian,utagi2020singularities,jagadish2020convex,megier2017eternal,wudarski2017robustness,shrikant2018non}.

\section{Design of Single Qubit Channels to Simulate}

Since we are only focused on single qubit channels, our channel of interest in this work will be the single qubit Pauli channel. These are the simplest non-trivial channels that we can use for our simulations. The single qubit Pauli channel is of the form,
\begin{equation}
\label{pauli_channel}
    \Lambda_{t}\rho=\sum_{\alpha=0}^{3}p_{\alpha}(t)\sigma_{\alpha}\rho\sigma_{\alpha},
\end{equation}
where $\sigma_{0}=\mathbb{1}$ and $p_{\alpha}(t)$ is a time dependent probability distribution such that $p_{0}(0)=1$ and $p_{i}(0)=0$ for $i \in \{1,2,3\}$ and $\sum_{i=0}^{3}p_{i}(t)=1$ for all $t \geq 0$.  This channel was studied extensively in \cite{vacchini2012classical}. It should also be noted that this channel also falls into a larger class of channels called random unitary channels \cite{chruscinski2015non}. In this work, we consider a subset of single qubit channels by focusing on single qubit Pauli channels; the reason we consider these channels is that the Pauli channel in equation (\ref{pauli_channel}) is the most general unital stochastic single qubit quantum channel \cite{vacchini2012classical,king2001minimal}. This means that it is the most general single qubit channel that preserves the identity operator and the most general single qubit channel with a Kraus operator, which is proportional to the identity; this is what it means for a channel to be a stochastic channel \cite{graydon2022designing}. Also, since the space of stochastic channels is manifestly convex \cite{graydon2022designing}, taking convex mixtures of single qubit Pauli channels will again be a stochastic channel. Therefore, simulating efficiently on a NISQ device, a single qubit Pauli channel provides insight into a broad class of single qubit channels, which is why we only consider single qubit Pauli channels.

To make a statement about whether the Pauli channel is Markovian or not, we need to analyze the decay rates of the time local generator $\mathcal{L}(t)$ of the Pauli channel in equation (\ref{pauli_channel}). From the time-local master equation (\ref{time_local_ME}), we see that $\mathcal{L}(t)=\dot{\Lambda}_{t}\Lambda_{t}^{-1}$, this tells us that we need to compute $\Lambda_{t}^{-1}$ to calculate the generator \cite{chruscinski2013non}. Let us note that:
\begin{equation}
    \Lambda_{t}(\sigma_{\alpha})=\lambda_{\alpha}(t)\sigma_{\alpha}
\end{equation}
where the time-dependent eigenvalues are,
\begin{equation}
    \lambda_{\alpha}(t)=\sum_{\beta=0}^{3}H_{\alpha \beta}p_{\beta}(t)
\end{equation}
with $\lambda_{0}(t)=1$ and $H_{\alpha \beta}$ being the Hadamard matrix defined as:
\begin{equation}
    H=\begin{pmatrix}
    1 & 1 & 1 & 1\\
    1 & 1 & -1 & -1\\
    1 & -1 & 1 & -1\\
    1 & -1 & -1 & 1\\
    \end{pmatrix}.
\end{equation}
We should note that $\lambda_{0}(t)=1$ and $|\lambda_{k}(t)| \leq 1$ for $k=1,2,3$. Now it is clear that:
\begin{equation}
    \mathcal{L}(t)\sigma_{\alpha}= \mu_{\alpha}(t) \sigma_{\alpha}
\end{equation}
where $\mu_{\alpha}(t)=\frac{\dot{\lambda}_{\alpha}(t)}{\lambda_{\alpha}(t)}$ and in particular $\mu_{0}(t)=0$ since $\lambda_{0}(t)=1$. Now by introducing the decay rate $\gamma_{\alpha}(t)$ we get the local generator as:
\begin{equation}
    \mathcal{L}(t)\rho=\sum_{\alpha=0}^{3}\gamma_{\alpha}(t)\sigma_{\alpha}\rho\sigma_{\alpha} 
\end{equation}
where the decay rates $\gamma_{\alpha}(t)$ are defined as,
\begin{equation}
\gamma_{\alpha}(t)=\frac{1}{4}\sum_{\beta=0}^{3}H_{\alpha \beta}\mu_{\beta}(t) .
\end{equation}
By observing that $\sum_{\alpha=0}^{3}\gamma_{\alpha}=0$, we arrive at the standard form for the generator of the Pauli channel \cite{chruscinski2013non} as,
\begin{equation}
\label{generator}
    \mathcal{L}(t)\rho=\sum_{k=1}^{3}\gamma_{k}(t) (\sigma_{k}\rho\sigma_{k} -\rho).
\end{equation}
Hence we get the expression for $\gamma_{k}(t)$ as:
\begin{equation}
\label{decay_rates}
    \gamma_{k}(t)=\frac{1}{4}\sum_{\beta=0}^{3}H_{k \beta} \Bigg\{ \frac{\sum_{\nu=0}^{3}H_{\beta \nu}\dot{p}_{\nu}(t)}{\sum_{\sigma=0}^{3}H_{\beta \sigma}p_{\sigma}(t)} \Bigg\},
\end{equation}
for $k=1,2,3$. We use the following result from \cite{chruscinski2013non,chruscinski2015non,vacchini2012classical}. The Pauli channel (\ref{pauli_channel}) is Markovian (i.e $\Lambda_{t}$ is divisible) if and only if $\gamma_{k}(t) \geq 0$ for all $t \geq 0$ and for $k=1,2,3$. This tells us that if the decay rates of the generator of a channel are non-negative, then that channel is Markovian, and any deviation from this leads to the channel becoming non-Markovian.
We know that from the time local generator $\mathcal{L}(t)$ we can get the channel $\Lambda_{t}$ from the relation,
\begin{equation}
    \Lambda_{t}=\mathcal{T}e^{\int_{0}^{t} d\tau \mathcal{L}(\tau)}
\end{equation}
where $\mathcal{T}$ is the chronological time ordering operator. Now we note that a linear combinations of Markovian generators $\mathcal{L}_{1}(t)$ and $\mathcal{L}_{2}(t)$ i.e. $\alpha_{1}\mathcal{L}_{1}(t)+\alpha_{2}\mathcal{L}_{2}(t) $  with $\alpha_{1},\alpha_{2} \geq 0$, is also a Markovian generator. Hence Markovian generators form a convex set in the space of admissible generators. The same is not true on the channel level, since for two Markovian channels $\Lambda_{t}^{(1)}$ and $\Lambda_{t}^{(2)}$ it is not always the case that their convex linear combination will be Markovian, i.e. $\eta\Lambda_{t}^{(1)}+(1-\eta)\Lambda_{t}^{(2)}$, $\eta \in [0,1]$, is not necessarily Markovian \cite{wudarski2016markovian}. We can now consider the two cases of convex mixing of channels outlined above, i.e. M+M=nM and nM+nM=M. We shall find Pauli channels that satisfy these two cases.

\subsection{Markovian Channel Addition (M+M=nM)}

The goal is to find two Markovian channels that, when convexly combined, yield a non-Markovian channel. We shall use the channels from \cite{uriri2020experimental} that demonstrate this. We start by defining the following two channels,
\begin{equation}
\label{lam_1_Markov}
    \Lambda_{t}^{(1)}\rho= p(t)\rho + (1-p(t))\sigma_{1}\rho\sigma_{1}
\end{equation}
\begin{equation}
\label{lam_2_Markov}
    \Lambda_{t}^{(2)}\rho= p(t)\rho + (1-p(t))\sigma_{2}\rho\sigma_{2}
\end{equation}
where the probability $p(t)=\frac{1+e^{-t}}{2}$. Using equations (\ref{decay_rates}) and (\ref{generator}) above, we can find the generators of these channels. We find that for channels $\Lambda_{t}^{(1)}=e^{t\mathcal{L}_{1}}$ and $\Lambda_{t}^{(2)}=e^{t\mathcal{L}_{2}}$ the generators are given by:

\begin{equation}
    \mathcal{L}_{1}\rho=\frac{1}{2}(\sigma_{1}\rho\sigma_{1}-\rho), \hspace{10mm} \mathcal{L}_{2}\rho=\frac{1}{2}(\sigma_{2}\rho\sigma_{2}-\rho).
\end{equation}
Since the decay rates in both these generators are non-negative for all $t \geq 0$, the channels in equations (\ref{lam_1_Markov}) and (\ref{lam_2_Markov}) are Markovian. We can now consider the convex combination of these channels in the following way:
\begin{align}
    \Lambda_{t}^{(T)}\rho&=\frac{1}{2}\Lambda_{t}^{(1)}\rho + \frac{1}{2}\Lambda_{t}^{(2)}\rho = \frac{1+e^{-t}}{2}\rho + \frac{1-e^{-t}}{4}(\sigma_{1}\rho\sigma_{1}+\sigma_{2}\rho\sigma_{2})
\end{align}
where $\Lambda_{t}^{(T)}$ is the total channel. Now for this total channel, the generator is:
\begin{align}
\label{lam_T_non_markov_generator}
    &\mathcal{L}_{T}(t)\rho=\frac{1}{4}(\sigma_{1}\rho\sigma_{1}-\rho)+\frac{1}{4}(\sigma_{2}\rho\sigma_{2}-\rho)-\frac{\mathrm{tanh}(t/2)}{4}(\sigma_{3}\rho\sigma_{3}-\rho)
\end{align}
Since the decay rate $\gamma_{3}(t) < 0 $ in equation (\ref{lam_T_non_markov_generator}), the total channel is non-Markovian more so this channel is eternally non-Markovian \cite{chruscinski2013non,wudarski2016markovian}.

\subsection{Non-Markovian Channel Addition (nM+nM=M)}

To design channels for the non-Markovian channel addition, we need to consider a special case of the Pauli channel in equation (\ref{pauli_channel}), i.e. the depolarizing channel \cite{garcia2020ibm}. We can obtain the depolarizing channel from equation (\ref{pauli_channel}) by parameterizing the probabilities $p_{\alpha}(t)$ as follows:
\begin{equation}
\label{depolarising_channel}
    \Lambda_{t}\rho=(1-\frac{3p(t)}{4})\rho + \frac{p(t)}{4}\sum_{\alpha=1}^{3}\sigma_{\alpha}\rho\sigma_{\alpha}
\end{equation}
where $0 \leq p(t) \leq 1$ for all times $t \geq 0$. Now from the fact that $\Lambda_{0}=\mathbb{1}$ we see that $p(0)=0$. The decay rates for the depolarizing channel can now be calculated using equation (\ref{decay_rates}),
\begin{equation}
\label{depolar_decay_rate}
    \gamma_{k}(t)=\frac{\dot{p}(t)}{4(1-p(t))}
\end{equation}
for k=1,2,3. From equation (\ref{depolar_decay_rate}), we can tighten the bounds on $p(t)$, i.e. $0 \leq p(t) < 1$. Now the form of the decay rate in equation (\ref{depolar_decay_rate}) tells us that for the channel to be Markovian, the function $p(t) $ should satisfy $\dot{p}(t) \geq 0 $ for all $t \geq 0$ and for the channel to be non-Markovian there should exist some time $t' \geq 0$ such that $\dot{p}(t') \leq 0$.\\
Now let us consider the following two individual channels:
\begin{align}
\label{non_markov_lam}
    &\Lambda_{t}^{(1)}\rho =(1-\frac{3q(t)}{4})\rho + \frac{q(t)}{4}\sum_{\alpha=1}^{3}\sigma_{\alpha}\rho\sigma_{\alpha}\nonumber \\
    &
    \Lambda_{t}^{(2)}\rho =(1-\frac{3r(t)}{4})\rho + \frac{r(t)}{4}\sum_{\alpha=1}^{3}\sigma_{\alpha}\rho\sigma_{\alpha}.
\end{align}
The decay rates for both the individual channels in equation (\ref{non_markov_lam}) are:
\begin{equation}
\label{decay_rates_non_markov_channels}
     \gamma_{k}^{(1)}(t)=\frac{\dot{q}(t)}{4(1-q(t))} \hspace{5mm}  ,\gamma_{k}^{(2)}(t)=\frac{\dot{r}(t)}{4(1-r(t))}
\end{equation}
for k=1,2,3. Taking a convex combination of the individual channels in equation (\ref{non_markov_lam}), we obtain the total channel:
\begin{align}
\label{total_markov_channel}
    \Lambda_{t}^{(T)}\rho&= \eta\Lambda_{t}^{(1)}\rho + (1-\eta)\Lambda_{t}^{(2)}\rho \nonumber\\
    &
    =(1-\frac{3}{4}w(t))\rho + \frac{w(t)}{4}\sum_{\alpha=1}^{3}\sigma_{\alpha}\rho\sigma_{\alpha}
\end{align}
where $\eta \in [0,1]$ and $w(t)=\eta q(t)+(1-\eta)r(t)$. Now the decay rates for the total channel are:\\
\begin{equation}
\label{decay_rate_total_Markov}
    \gamma_{k}^{(T)}(t)=\frac{\dot{w}(t)}{4(1-w(t))}=\frac{\eta\dot{q}(t)+(1-\eta)\dot{r}(t)}{4(1-[\eta q(t)+(1-\eta)r(t)])}
\end{equation}
It is clear from equation (\ref{decay_rate_total_Markov}) that it is possible for there to exist times $t'$ and $t''$ such that $\dot{q}(t')<0$ and $\dot{r}(t'')<0$ and $w(t) \geq 0$ for all times $t$. This tells us that it is possible to use the depolarizing channel in equation (\ref{depolarising_channel}) to show that we can convexly mix two non-Markovian channels to get a total Markovian channel. 

As a consequence of our calculations we can also see that it is impossible to have mixtures of two Markovian depolarising channels that yield a Non-Markovian depolarising channel. To see this consider the decay rates in equation (\ref{decay_rates_non_markov_channels}), for the channels $\Lambda_{t}^{(1)}$ and $\Lambda_{t}^{(2)}$ to be markovian we require that $\dot{q}(t)\geq 0$ and $\dot{r}(t)\geq 0$ for all times $t\geq 0$. However if this is true then the decay rates for total channel $\Lambda_{t}^{(T)}$ in equation (\ref{decay_rate_total_Markov}) will always be positive since, $\eta\dot{q}(t)+(1-\eta)\dot{r}(t) \geq 0$ for all $t\geq 0$. This tells us that mixing two Markovian depolarising channels always yields a Markovian depolarising channel. This result is useful as depolarising channels are the most commonly used noise models when studying the robustness of quantum algorithms \cite{pillay2023multi,cross2015quantum} to noise.

The goal now would be to pick functions $q(t)$ and $r(t)$ such that $\Lambda_{t}^{(1)}$ and $\Lambda_{t}^{(2)}$ are non-Markovian and the total channel $\Lambda_{t}^{(T)}$ is Markovian. We observe that if we parameterize $q(t)$ and $r(t)$ as:
\begin{figure}[h!]
  \centering
  \includegraphics[width=0.47\textwidth]{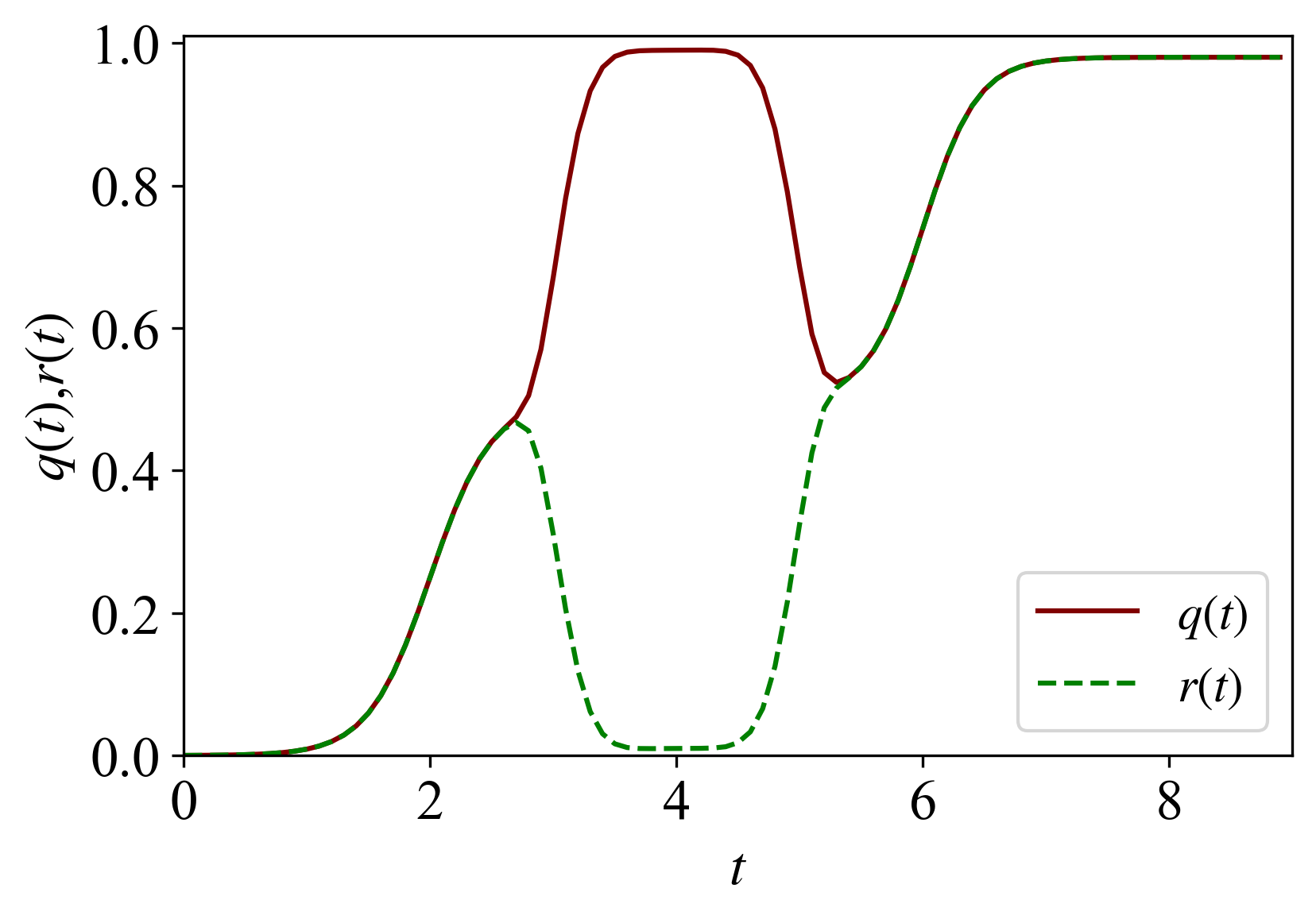}
  \caption{This is a plot of the functions $q(t)$ and $r(t)$ for the time interval $t \in [0,9]$. Both these functions satisfy the constraints that $0 \leq q(t), r(t) <1$ and the functions are numerically close enough to zero that we can write: $q(0) \approx 0$ and $r(0) \approx 0$.}
  \label{QandR}
  \end{figure}
  
\begin{align}
    &q(t)=a(t)+b(t)\nonumber\\
   \nonumber\\
    &
    r(t)=a(t)-b(t)
\end{align}
Furthermore, setting $\eta=\frac{1}{2}$ in equation (\ref{total_markov_channel}), then using the bounds on $q(t)$ and $r(t)$, we get $b(t) \leq a(t) < 1-b(t) $. Taking into consideration all the constraints outlined above, we choose the functions $a(t)$ and $b(t)$ as follows:
\begin{align}
\label{functions_a_b}
    &a(t)=\frac{0.5}{1+\mathrm{exp}(-4(t-2))}+\frac{0.48}{1+\mathrm{exp}(-4.5(t-6))}\nonumber\\
    \nonumber\\
    &
    b(t)=0.49\mathrm{exp}(-(t-4)^{6}).
\end{align}
Refer to Fig. \ref{QandR} for the plots of the functions given in equation (\ref{functions_a_b}) above. From equation (\ref{decay_rates_non_markov_channels}), we see that the decay rates for the individual channels $\Lambda_{t}^{(1)}$ and $\Lambda_{t}^{(2)}$ are non-Markovian. This is because for some time interval $\dot{q} <0$ and for some other interval of time $\dot{r}<0$ making the decay rates negative, leading to the channels being non-Markovian. Fig. \ref{QdandRd} shows the plots of $\dot{q}(t)$ and $\dot{r}(t)$, which shows that both are negative for some times. The total channel $\Lambda_{t}^{(T)}$ is parameterized by the function $w(t)$ and it is clear from Fig. \ref{Fig3} that $\dot{w}(t) \geq 0$ for all times $t\geq 0$, so the total channel $\Lambda_{t}^{(T)}$ is Markovian. Refer to Appendix A for more intuition about how the functions $q(t)$ and $r(t)$ were chosen. Hence we have found an example of the convex sum of two non-Markovian channels $\Lambda_{t}^{(1)}$ and $\Lambda_{t}^{(2)}$  yielding a Markovian total channel $\Lambda_{t}^{(T)}$.\\

\begin{figure}[h!]
  \centering
  \includegraphics[width=0.47\textwidth]{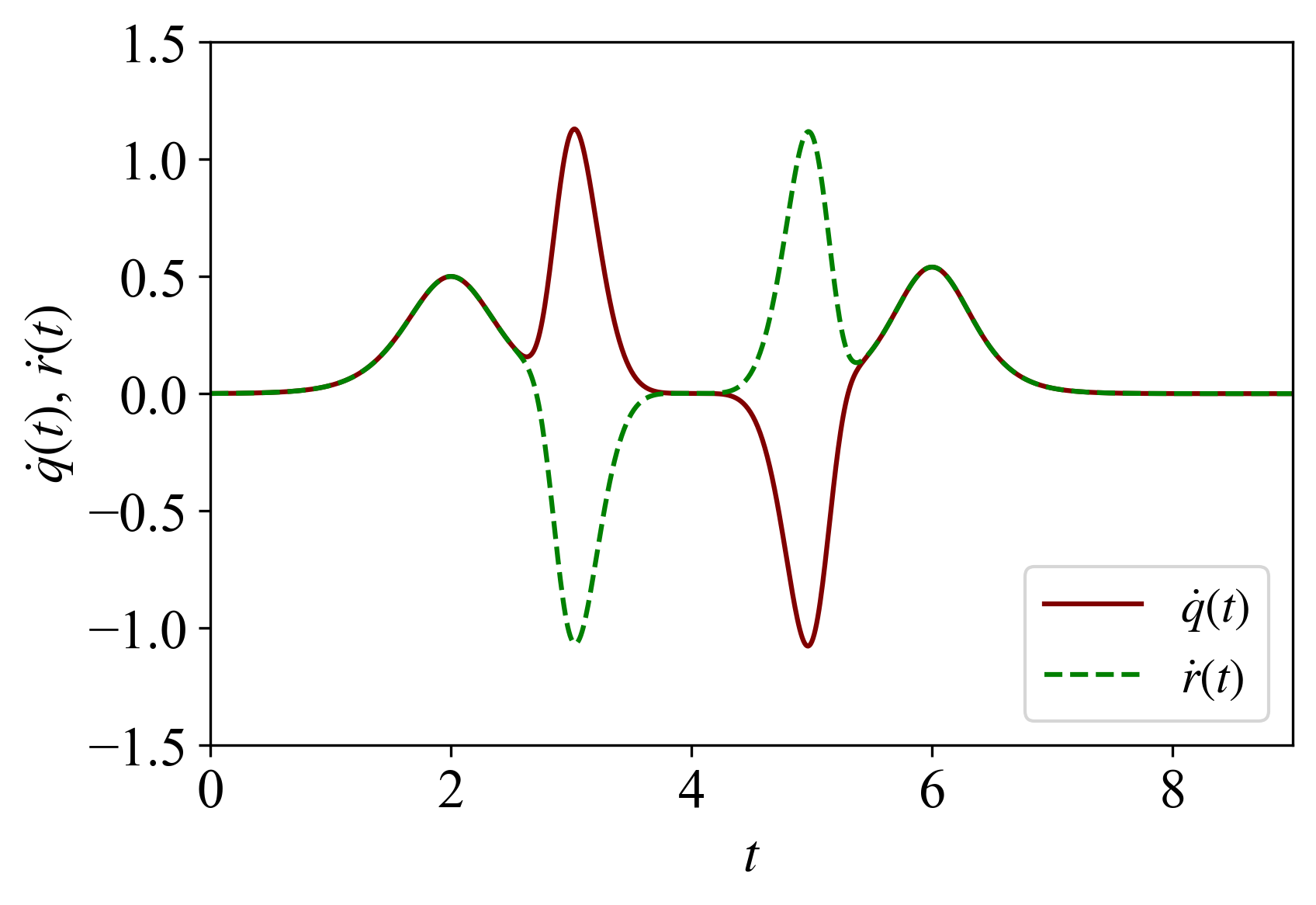}
  \caption{This plot shows the derivatives of $q(t)$ and $r(t)$, we see that for some times both $\dot{q}(t)$ and $\dot{r}(t)$ are negative which will lead to the decay rates in the generators of the individual channels to become negative. This implies that the individual channels are both non-Markovian.}
  \label{QdandRd}
  \end{figure}
  
\begin{figure}[h!]
  \centering
  \includegraphics[width=0.47\textwidth]{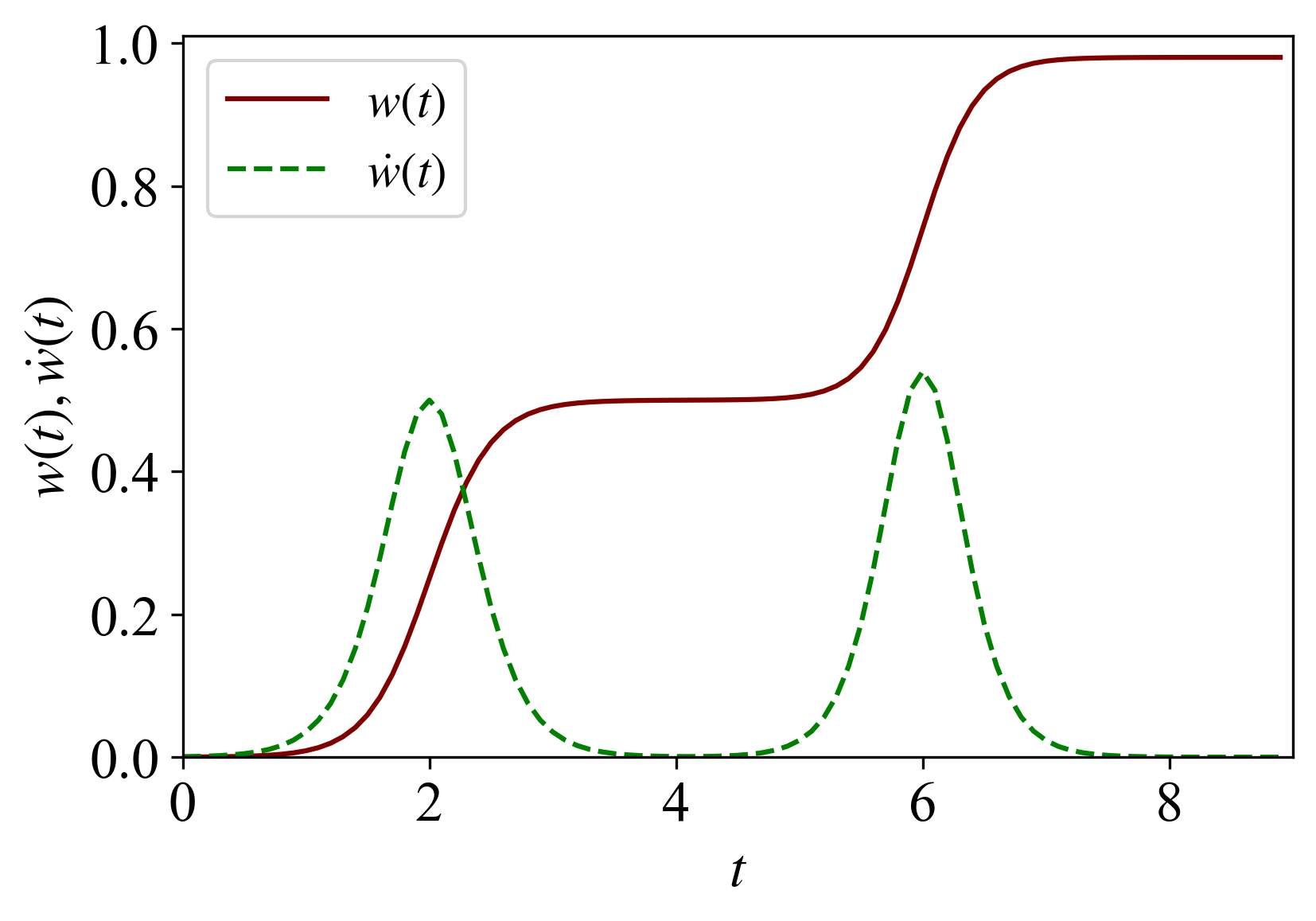}
  \caption{Plot showing the function $w(t)$ which parameterizes the total channel $\Lambda_{t}^{(T)}$, one can easily see that $w(t)$ satisfies all the constraints above. This plot also shows $\dot{w}(t)$, and it is clear that $\dot{w}(t) \geq 0$ for all times $t$, which tells us that the decay rates of the total channel are all greater than or equal to zero which implies that the total channel $\Lambda_{t}^{(T)}$ is Markovian.}
  \label{Fig3}
\end{figure}

\section{Simulation of Quantum Channels on a NISQ Device}

The simulation of the channels from the previous section on a NISQ device requires a twofold method. First, one needs to construct quantum circuits that implement the action of the quantum channel on a single qubit. Then we should perform a quantum process tomography (QPT) to reconstruct the quantum channel. The tomography will yield a non-physical quantum channel. Hence convex optimization must be used to construct the closest possible physical quantum channel to the channel obtained from the QPT. The following sub-sections shall detail the method above.

\subsection{Quantum Circuits for simulation of the Channels on a NISQ device}

To simulate the quantum channels in section 3, we must construct a quantum circuit that implements the channel. Usually, this is done by naively using the Stinespring representation of the channel \cite{stinespring1955positive}, but this naive application of the Stinespring dilation may lead to a complicated quantum circuit unsuitable for the NISQ setting. To solve this problem we use the Stinespring representation in a way that takes into account the properties of the NISQ device, when we construct circuits for the channels. To design quantum circuits for our channels from section 2 that are appropriate for the NISQ setting, we take advantage of the fact that in both cases (i.e. M+M=nM and nM+nM=M), the channels that were designed implement a Pauli operator with some probability. We can easily design circuits that can implement the channels by using ancilla qubits. This approach is inspired by \cite{garcia2020ibm} and allows us to construct NISQ-appropriate quantum circuits.

\subsubsection{Circuits for Markovian Channel addition}

To implement the Markovian channel $\Lambda_{t}^{(1)}$ with probability $p(t)$ we use the circuit shown in Fig. \ref{fig:circ_1} below. In the circuit the gate $R_{y}(\theta)$ is a rotation gate and its matrix representation is,
\begin{align}
    R_{y}(\theta)=\begin{pmatrix}
        \cos(\frac{\theta}{2}) & & -\sin(\frac{\theta}{2})\\
        \\
       \sin(\frac{\theta}{2}) & & \cos(\frac{\theta}{2})
    \end{pmatrix},
\end{align}
where $\theta$ is the angle of rotation. The circuit is understood as follows it will apply the gate $\sigma_{1}=X$ to the input state $\rho_{\mathrm{in}}$ with probability $|\sin(\theta/2)|^{2}$ and it will leave the state unchanged with probability $|\cos(\theta/2)|^{2}$ (i.e. when the ancilla is in the state $|0\rangle$ do not change the state $\rho_{\mathrm{in}}$ and when the ancilla is in state $|1\rangle$ apply the $X$ gate to the input state $\rho_{\mathrm{in}}$). From this, we can get the value of the angles in terms of the probability $p(t)$ as:
\begin{equation}
\label{thetaeq}
    \theta=2\mathrm{arccos}(\sqrt{p(t)})=2\mathrm{arccos}\bigg(\sqrt{\frac{1+e^{-t}}{2}} \hspace{1mm}\bigg)
\end{equation}
Since the angle $\theta$ is written in terms of $p(t)$ the circuit in Fig. \ref{fig:circ_1} will leave $\rho_{\mathrm{in}}$ unchanged with probability $p(t)$ and it will apply $\sigma_{1}$ to $\rho_{\mathrm{in}}$ with probability $1-p(t)$. Hence the circuit implements the channel $\Lambda_{t}^{(1)}$.
\begin{figure}[h]
   \centering
   \includegraphics[scale=0.2]{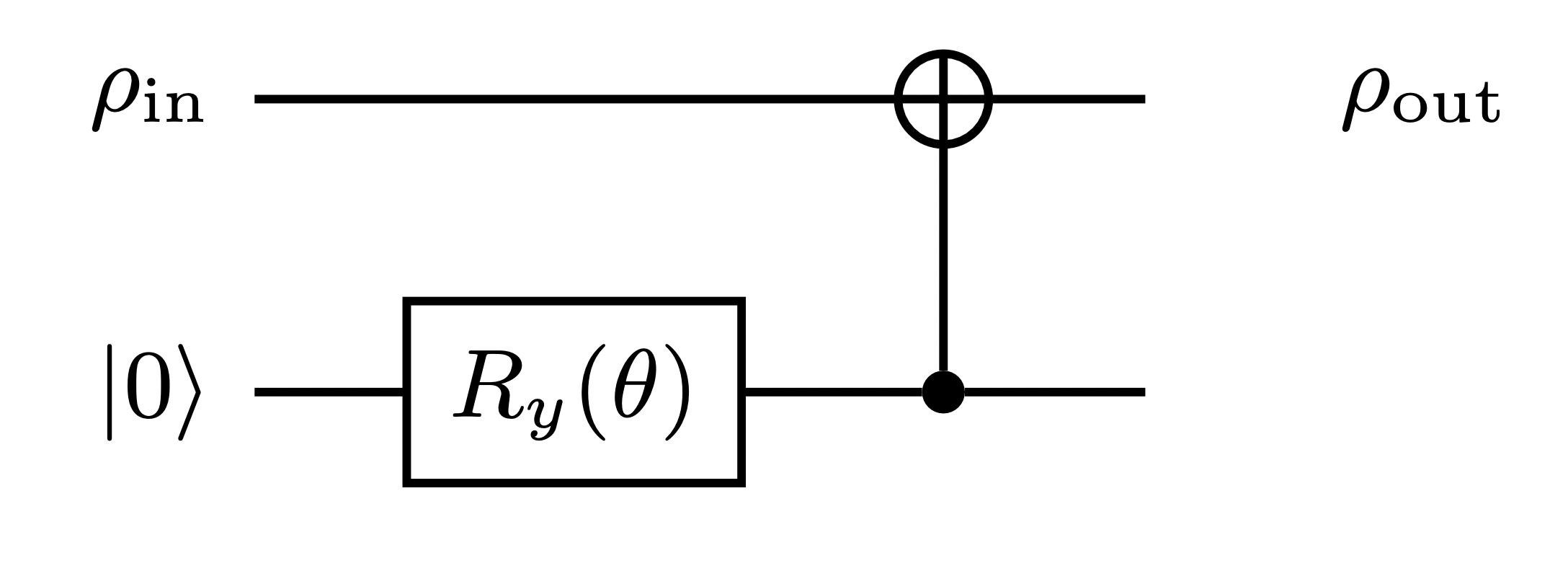}
   \caption{Quantum circuit implementing the Markovian channel $\Lambda^{(1)}_t$ for probability $p(t)$.}
   \label{fig:circ_1}
\end{figure}
Similarly, we can design a circuit that implements $\Lambda_{t}^{(2)}$ by using the same ideas used to design the previous circuit. Fig. \ref{fig:circ_2} shows the quantum circuit that implements the channel $\Lambda_{t}^{(2)}$ where the $Y$ gate is applied to the input state with probability $p(t)$. The angle $\theta$ is obtained using equation (\ref{thetaeq}).
Now that we have designed circuits that implement the individual Markovian channels, we design a circuit that implements the total non-Markovian channel $\Lambda_{t}^{(T)}$. In designing this circuit, we keep in mind that we want to leave the input state $\rho_{\mathrm{in}}$ unchanged with probability $p(t)$ and apply the noise operators $\sigma_{1}$ and $\sigma_{2}$ to the input state with probability $\frac{1-p(t)}{2}$. Refer to Fig. \ref{fig:circ_3} for the quantum circuit that implements the total channel. This circuit implements the channel by first preparing the two ancilla qubits in the state: $\cos(\theta_{1}/2)|00\rangle -\sin(\theta_{1}/2)\sin(\theta_{2}/2)|01\rangle -\sin(\theta_{1}/2)\cos(\theta_{2}/2)|11\rangle$ now using the ancilla qubits the circuit will apply $X$ to the input state if the ancillae are in the state $|01\rangle$, it will apply $Y$ to the input state when the ancillae are in the state $|11\rangle$ (NB: the circuit applies $XZ$ to the input state which is the same as applying $Y$ up to a phase). The circuit leaves the input state unchanged when the ancillas are in the state $|00\rangle$. From the state of the ancillae, we can determine the angles $\theta_{1}$ and $\theta_{2}$:
\begin{equation}
    \theta_{1}=2\mathrm{arccos}(\sqrt{p(t)}), \hspace{10mm} \theta_{2}=\frac{\pi}{2}
\end{equation}
Now that we have the angles in terms of $p(t)$, we have found the circuit that implements the channel $\Lambda_{t}^{(T)}$.
\begin{figure}[h]
   \centering
   \includegraphics[scale=0.2]{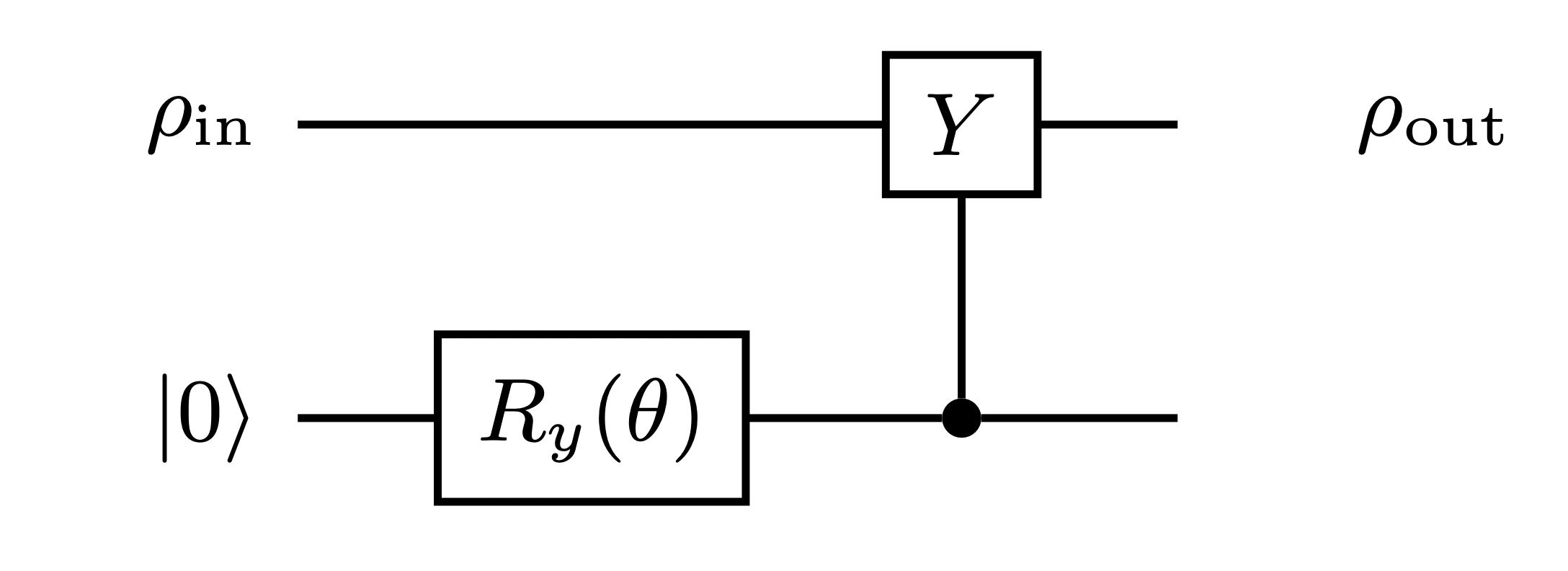}
   \caption{Quantum circuit implementing the Markovian channel $\Lambda^{(2)}_t$ for probability $p(t)$. }
   \label{fig:circ_2}
\end{figure}
\begin{figure}[h]
   \centering
  \includegraphics[scale=0.2]{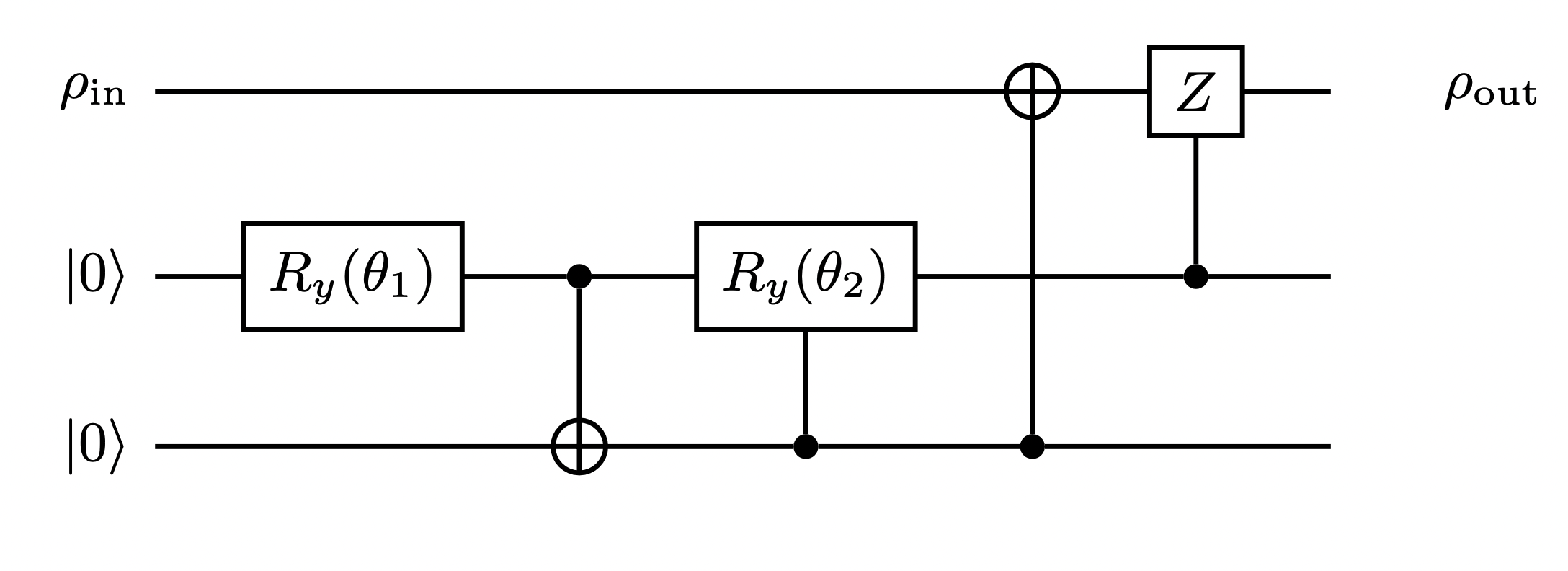}
   \caption{Quantum circuit implementing the total non-Markovian channel $\Lambda^{(T)}_t$ for probability $p(t)$.}
   \label{fig:circ_3}
\end{figure}
\subsubsection{Circuits for non-Markovian Channel addition}

\begin{figure}[h]
   \centering
   \includegraphics[scale=0.2]{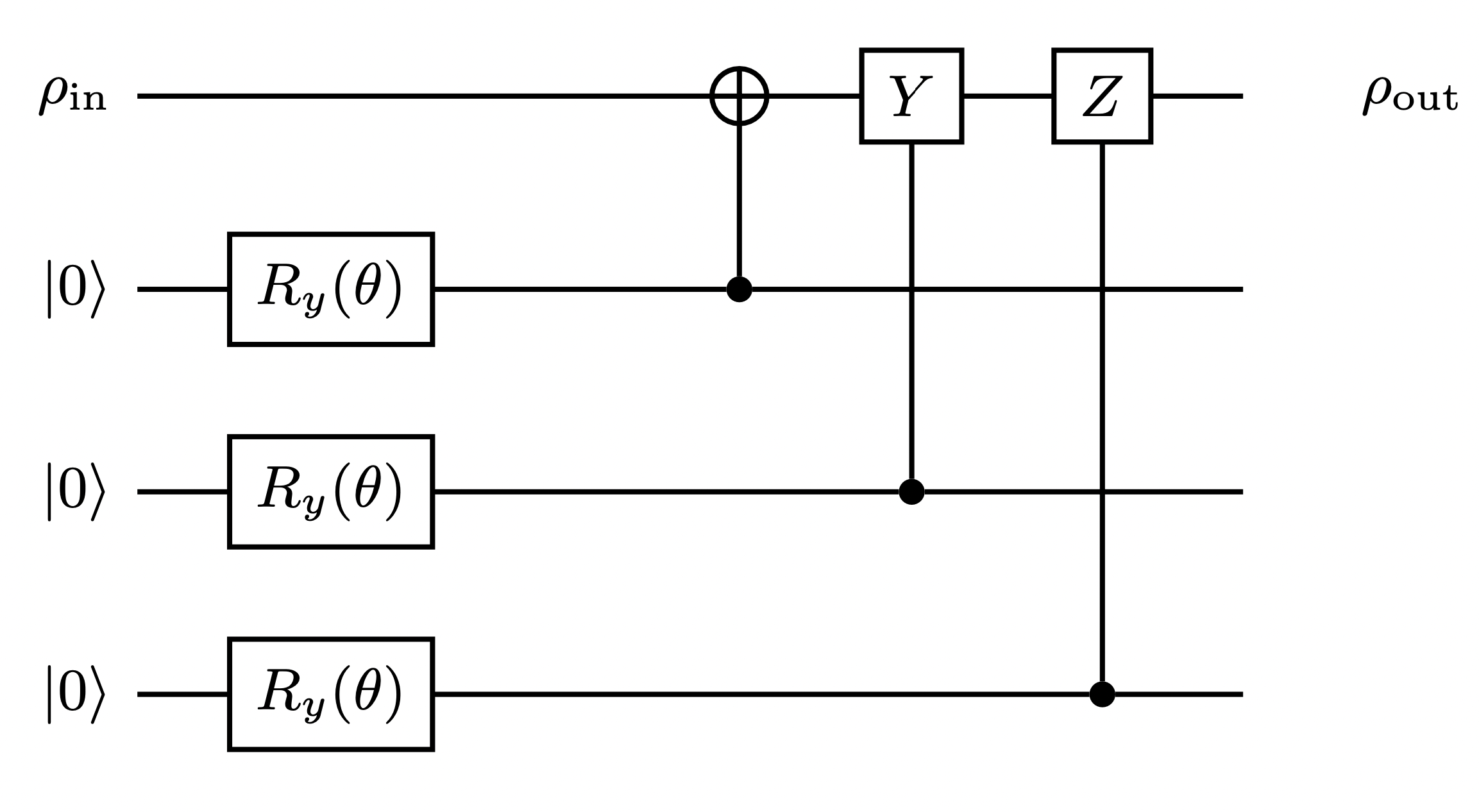}
   \caption{circuit implementing the depolarizing channel for a single system qubit, for the probability $p(t)$. The angle $\theta$ is determined by the formula $\theta(t)=\frac{1}{2}\mathrm{arccos}(1-2p(t))$.}
   \label{fig:circ_4}
\end{figure}

To implement the channels that we have designed for the non-Markovian channel addition, we need a circuit that can implement the depolarizing channel that is parameterized by some function $p(t)$ (Refer to equation (\ref{depolarising_channel})). We use the circuit simulating the depolarizing channel in \cite{garcia2020ibm} to do this. Refer to Fig. \ref{fig:circ_4} for the circuit that simulates the depolarising channel. The three ancillea qubits are prepared in the state $(\cos(\theta/2)\ket{0}+\sin(\theta/2)\ket{1})^{\otimes 3}$ are used as control qubits for the controlled-$X$, controlled-$Y$ and controlled-$Z$ gates, so that the gates $X,Y$ and $Z$ are all applied to the input state $\rho_{in}$ with probability $\sin^{2}(\theta/2)$. To use this circuit to implement the depolarising channel in equation (\ref{depolarising_channel}) the rotation angle $\theta$  needs to be in terms of the probability $p(t)$ so that each gate is applied with probability $p(t)/4$. We observe that applying $X$ and then $Z$ to $\rho_{in}$ but not $Y$, is equivalent to just applying $Y$ to $\rho_{in}$ and so on. The resulting equation that relates the probability $p(t)$ to $\theta$ is,
\begin{align}
    \sin^{2}(\frac{\theta}{2})\cos^{4}(\frac{\theta}{2})+\cos^{2}(\frac{\theta}{2})\sin^{4}(\frac{\theta}{2})=\frac{p(t)}{4}.
\end{align}
Solving this equation for $\theta$ in terms of $p(t)$ we get,
\begin{align}
    \theta(t)=\frac{1}{2}\arccos(1-2p(t)),
\end{align}
which will be the parameter in the $R_{y}$ gate in then circuit in Figure \ref{fig:circ_4}.
\subsection{Tomographic Reconstruction of the Channel}
 It is known that a quantum channel $\Lambda_{t}$ has a Kraus representation \cite{kraus1971general}:
\begin{equation}
    \Lambda_{t}\rho=\sum_{\alpha}\hat{K}_{\alpha}\rho\hat{K}_{\alpha}^{\dagger}
\end{equation}
where $\hat{K}_{\alpha}$ are the Kraus operators that satisfy $\sum_{\alpha}\hat{K}_{\alpha}^{\dagger}\hat{K}_{\alpha}=\mathbb{1}$. In this work, we will consider the case of a single qubit channel then the Kraus operators are $2\times2$ matrices. If we choose a complete basis for the Kraus operators of a single qubit channel as $\{ \sigma_{0}= \mathbb{1},\sigma_{1},\sigma_{2},\sigma_{3} \}$, where $\sigma_{i}$ are the usual Pauli matrices. Then we can expand the Kraus operators in terms of this basis to get the process matrix representation of the quantum channel for a single qubit:
\begin{equation}
\label{chi_representation}
    \Lambda_{t}\rho=\sum_{m,n=0}^{3} \chi_{mn}\sigma_{m}\rho\sigma_{n} .
\end{equation}
Here $\chi_{mn}$ is a positive and Hermitian $4\times4$ matrix called the process matrix and shall be determined using a quantum process tomography \cite{chuang1997prescription, NC10}. Now if we know the process matrix, then we have a complete description of the channel $\Lambda_{t}$.\\
To determine the elements of the process matrix, we need to choose a complete set of input states. We choose the states
\begin{align}
	\mathcal{D}&=\bigg\{\ket{\psi_{1}}=|0\rangle , \ket{\psi_{2}}=|1\rangle,\nonumber\\
	&\ket{\psi_{3}}=|+\rangle=\frac{1}{\sqrt{2}}(|0\rangle +|1\rangle),\nonumber\\
	&\ket{\psi_{4}}=|+_{y}\rangle=\frac{1}{\sqrt{2}}(|0\rangle +i|1\rangle)\bigg\}.
\end{align}
 The states from the set $\mathcal{D}$ form a complete set as the projectors constructed from each ket vector in this set can be used to construct the density operator of any physical single qubit state. The input states are sent through the channel, $\Lambda_{t}$. We can prepare the initial state of the qubit as each input state. Using quantum state tomography, we reconstruct the state after each input state is passed through the channel \cite{james2005measurement}. Then the formulas from \cite{chuang1997prescription,NC10} are used to construct the $\chi$ matrix which allows us to reconstruct the channel $\Lambda_{t}$.\\
 To perform a QPT on a quantum computer, we use the circuits from sections 3.1.1 and 3.1.2, and then the system qubit is prepared in one of the input states above and passed through the channel. We then perform the state tomography on the system qubit for each input state and get the corresponding counts. Using the counts obtained from the tomographic circuits, we can construct the process matrix \cite{chuang1997prescription}.

We use the counts obtained from the tomographic circuits to construct an initial process matrix denoted by, $\chi_{in}$. The matrix $\chi_{in}$ will not be positive and Hermitian. This is because we can only make a finite number of measurements on the system qubit. To correct this, we shall solve a convex optimization problem to find the closest possible process matrix, denoted $\chi_{\mathrm{c}}$, to the ideal process matrix, which we denote as $\chi_{\mathrm{id}}$  \cite{huang2020reconstruction}. We shall use the matrix $\chi_{in}$ as an initial condition when solving the optimization problem. To construct the optimization problem, we first need to parameterize $\chi_{c}$ in terms of parameters that we can optimize. We also define $\chi_{c}$ to be Hermitian and positive semi-definite for all parameter values. We now parameterise $\chi_{c}$ as,
 \begin{equation}
    \chi_{c}=\chi_{c}(x_{1},x_{2},...,x_{16})=T^{\dagger}T\end{equation}
where $T=T(x_{1},x_{2},...,x_{16})$ is a $4\times4$ triangular matrix that is a function of 16 real variables $x_{1},x_{2},...,x_{16}$ and is shown below in matrix form:
\begin{equation}
    T=\begin{pmatrix}
    x_{1} & 0 & 0 & 0 \\
    x_{5}+ix_{6} & x_{2} & 0 & 0\\
    x_{11}+ix_{12} &x_{7}+ix_{8} & x_{3} & 0\\
    x_{15}+ix_{16} & x_{13}+ix_{14} &x_{9}+ix_{10} & x_{4}\\
    \end{pmatrix}.
\end{equation}
It is evident from this parameterization that $\chi_{c}$ is positive semi-definite. Consider some arbitrary four-dimensional vector $\ket{\psi}$ then, 
\begin{align}
	\bra{\psi}T^{\dagger}T\ket{\psi} \geq 0.
\end{align}
It is also evident that $\chi_{c}$ is Hermitian.
To find $\chi_{c}$ with convex optimization, we need to define an objective function that will be minimized with respect to constraints. We define the objective function by the squared difference between the theoretical and experimental probability distributions for each of the counts obtained from the process tomography. The following projective measurement operators are defined from the set $\mathcal{D}$ above,

 \begin{align}
 	\left\{M_{1}=\ketbra{\psi_{1}}, M_{2}=\ketbra{\psi_{2}},M_{3}=\ketbra{\psi_{3}}, M_{4}=\ketbra{\psi_{4}} \right\}.
 \end{align}

 Next, we consider the input state $\rho_{i}=\ketbra{\psi_{i}}$ where $\ket{\psi_{i}} \in \mathcal{D}$. Then the theoretical probability of being in the state $\ket{\psi_{j}}$ after the application of the channel $\Lambda_{t}$ to the initial state $\ket{\psi_{i}}$, denoted $p_{ij}^{\text{the}}$, is,
 \begin{align}
 	p_{ij}^{\text{the}}&=\mathrm{Tr}[M_{j}\Lambda_{t}(\rho_{i})]\\
 	&=\mathrm{Tr}\left[M_{j}\left(\sum_{m,n=1}^{4}(\chi_{c})_{mn}\sigma_{m}\rho_{i}\sigma_{n}\right)\right]\\
 	&=\sum_{m,n=1}^{4}(\chi_{c})_{mn}\mathrm{Tr}\left[M_{j}\sigma_{m}\rho_{i}\sigma_{n} \right].
 \end{align}
The experimentally obtained probability $p_{ij}^{\text{exp}}$ of being the initial state $\rho_{i}$ and measuring in the state $\ket{\psi_{j}}$ is,
\begin{align}
	p_{ij}^{\text{exp}}=\frac{n_{ij}}{N},
\end{align}
where $n_{ij}$ is the counts obtained from the circuit with input state $\ket{\psi_{i}}$ and output state $\ket{\psi_{j}}$ and $N$ being the total number of counts. 
Now we can define the objective function as,
\begin{align}
	&\mathcal{F}(x_{1},...,x_{16})=\sum_{i,j=1}^{4}\left(p_{ij}^{\text{exp}}-p_{ij}^{\text{the}} \right)^{2}\\
	&=\sum_{i,j=1}^{4}\left(\frac{n_{ij}}{N}-\sum_{m,n=1}^{4}\left(\chi_{c}\right)_{mn}\mathrm{Tr}\left[M_{j}\sigma_{m}\rho_{i}\sigma_{n} \right]\right)^{2}.
\end{align}
We should also consider that the channel $\Lambda_{t}$ should be trace-preserving. This will be one of the constraints we define for our optimization problem. From equation (\ref{chi_representation}) we see that for $\Lambda_{t}$ to be trace-preserving we must have that,
\begin{align}
\sum_{m,n=1}^{4}(\chi_{c})_{mn}\sigma_{n}\sigma_{m}=\mathbb{1}.
\end{align}
We can weaken this constraint by just requiring that $\Lambda_{t}$ be trace non-increasing, this leads to,
\begin{align}
	\sum_{m,n=1}^{4}(\chi_{c})_{mn}\sigma_{n}\sigma_{m}\leq \mathbb{1}.
\end{align}
This constraint can be written as a positive semi-definite constraint i.e.
\begin{align}
	\mathbb{1}-\sum_{m,n=1}^{4}(\chi_{c})_{mn}\sigma_{n}\sigma_{m} \geq 0.
\end{align}
We can now state the optimization problem that we want to solve that will yield a positive semi-definite and Hermitian matrix $\chi_{c}$ that is the closest to $\chi_{\mathrm{id}}$,
\begin{align}
	\min_{\{x_{1},...,x_{16}\}} &\ \mathcal{F}(x_{1},...,x_{16}) \\
	\text{such that,} & \ \left(\mathbb{1}-\sum_{m,n=1}^{4}(\chi_{c})_{mn}\sigma_{n}\sigma_{m} \right)\geq 0, \\
	 & \chi_{c}(x_{1},...,x_{16})\geq 0,
\end{align}
where we obtain an initial set of parameters $(x^{(in)}_{1},...,x^{(in)}_{16})$ from $\chi_{in}$. The solution to this problem will yield the optimal values for $x_{1},...,x_{16}$ so that $\chi_{c}$ is as close to $\chi_{\text{id}}$ as possible. The following section will use the matrix $\chi_{c}$ for the characterization.

\section{Characterization of the Channel as Markovian or Non-Markovian}

From the previous section, we know that the state of the system at some time $t \geq 0$ is given in terms of the quantum channel $\Lambda_{t}$ as, $\rho(t)=\Lambda_{t}\rho(0)$, where $\rho (0)$ is the initial state of the system at $t=0$. We shall use the CP divisibility criteria to characterize the channel $\Lambda_{t}$ as Markovian or non-Markovian. We start by writing the dynamical map $\Lambda_{t}$ in the following way:
\begin{equation}
\label{intermediate_map}
    \Lambda_{t}=V_{t,s}\Lambda_{s}
\end{equation}
where $V_{t,s}$ is called the intermediate map (IM) from time $s$ to $t$. The maps $V_{t,s}$ form a family of two-parameter propagators. We say that $\Lambda_{t}$ is CP divisible if $V_{t,s}$ is completely positive (CP) for all $t \geq s \geq 0$. The goal is to check that the map $V_{t,s}$ is CP. We will use the techniques in \cite{uriri2020experimental}. From Section 3, we know that we can express a quantum channel in the following representation:
\begin{equation}
    \Lambda_{t}\rho=\sum_{m,n=0}^{3} \chi_{mn}\sigma_{m}\rho\sigma_{n},
\end{equation}
where $\chi_{mn}$ is a positive and Hermitian $4\times4$ matrix called the chi matrix, $\sigma_{0}=\mathbb{1}$ and $\sigma_{i}$ for $i=1,2,3$ are the Pauli matrices.  Knowing the chi matrices for two different time durations, $s$ and $t$, we can check whether the map $V_{t,s}$ that evolves the system from time $s$ to time $t$ is completely positive. To check this, we make use of the transfer matrices $F(s)$ and $F(t)$ for the maps $\Lambda_{s}$ and $\Lambda_{t}$, respectively. The transfer matrix $F(t)$ of a map $\Lambda_{t}$ is a concrete matrix representation of the map in a given orthonormal basis \cite{wolf2012quantum}. The transfer-matrix approach is useful as it allows us to represent the density matrix $\rho$ as a stacked vector $|\rho\rangle\rangle$, now the evolution of the vector $|\rho(0)\rangle\rangle$ can be written as: $|\rho(t)\rangle\rangle = F(t)|\rho(0)\rangle\rangle$ which is nothing more than the action of the transfer matrix on the stacked vector \cite{uriri2020experimental}. The elements of a transfer matrix $F(t)$ are given explicitly as,

\begin{equation}
\label{transfer_mat_elements}
    F_{\alpha,\beta}(t)=\mathrm{Tr}[G_{\alpha}^{\dagger} \Lambda_{t}G_{\beta} ] 
\end{equation}
\\
where $\{ G_{\alpha}\}$ are a set of orthonormal operators with respect to the Hilbert-Schmidt inner product \cite{wolf2012quantum}. We choose the set $\{ G_{\alpha}\}$ to be the standard matrix basis of $\mathcal{M}_{2}(\mathbb{C})$, i.e. $\{G_{1}=|0\rangle\langle 0|, G_{2}=|0\rangle\langle 1|, G_{3}=|1\rangle\langle 0|, G_{4}=|1\rangle\langle 1| \}$, where $\{|0\rangle,|1\rangle \}$ are the standard computational basis vectors of the single qubit. If we know the $\chi$ matrix for some time $t$, we can use this to calculate $\Lambda_{t}G_{\beta}$ and hence calculate $F(t)$. Now using equation (\ref{intermediate_map}) and writing it in terms of transfer matrices, we arrive at $F(t)=F(t,s)F(s)$. This tells us that if we have the transfer matrix for two times $s$ and $t$, we can get the transfer matrix of the intermediate map, i.e. $F(t,s)=F(t)F^{-1}(s)$. For a given transfer matrix $F(t)$ we can obtain the Choi matrix $W(t)$. For a single qubit, this can be written as:
\begin{equation}
\label{Choi_mat}
    W(t)= \frac{1}{2} \sum_{\alpha ,\beta=1}^{4} F_{\alpha ,\beta}(t)( G_{\beta} \otimes G_{\alpha})
\end{equation}

This is derived by applying $\Lambda_{t}$ to a single qubit of the maximally entangled state $|\beta_{00}\rangle =\frac{1}{\sqrt{2}}(|00\rangle +|11\rangle)$, hence $W(t)=(\mathbb{1} \otimes \Lambda_{t})|\beta_{00}\rangle \langle \beta_{00}|$. By the Choi-Jamiolkowski isomorphism, a dynamical map $\Lambda_{t}$ is completely positive if and only if the corresponding Choi matrix of the map is positive \cite{choi1975completely,jamiolkowski1972linear}. This tells us that the map $\Lambda_{t}$ is CP if all the eigenvalues of the Choi matrix $W(t)$ are non-negative. So for two times $s$ and $t$ such that $t \geq s \geq 0$, if the eigenvalues of the Choi matrix $W(t,s)$, of the intermediate map $V_{t,s}$, are non-negative then the intermediate map is completely positive and by our definition of CP divisibility, this tells us that the dynamical map $\Lambda_{t}=V_{t,s}\Lambda_{s}$ is Markovian. Any deviation from this leads to non-Markovian dynamics.

\section{Results and Discussion}

\begin{figure}[h]
    \centering
    \includegraphics[scale=0.3]{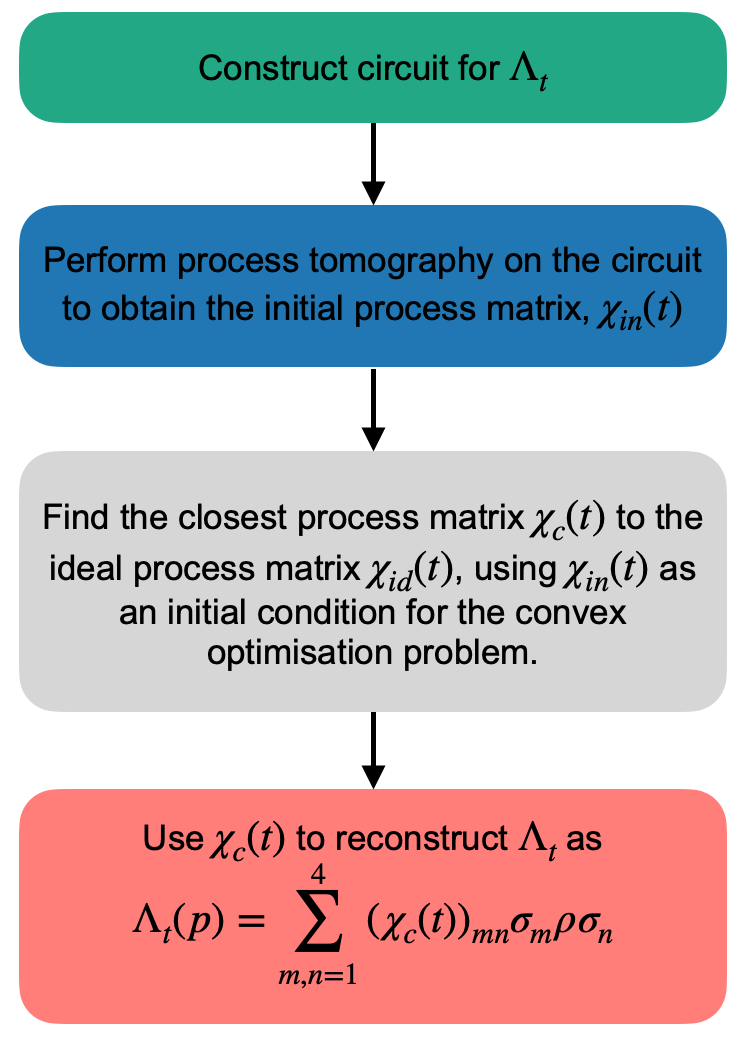}
    \caption{A flow chart summarising the experimental procedure used in the simulation and reconstruction of the channels for both cases M+M=nM and nM+nM=M.}
    \label{Fig20}
\end{figure}

This research aimed to demonstrate that more complex open quantum systems can be efficiently simulated on NISQ devices. In particular, we aimed to show that we can simulate convex mixtures of Markovian and Non-Markovian quantum channels on NISQ devices. In this section, we present quantum circuits based on the circuits in Section 4.2 that account for the topology of the NISQ devices they will be run on. We also present the results of the simulations and evaluate the quality of the results using three metrics: fidelity of the process matrix (discussed in Section 6.2), trace norm of the Choi matrix (discussed in Section 6.3) and the minimum eigenvalue of $W(t,s)$ in Section 6.4.

\begin{figure}[h]
    \centering
    \includegraphics[scale=0.3]{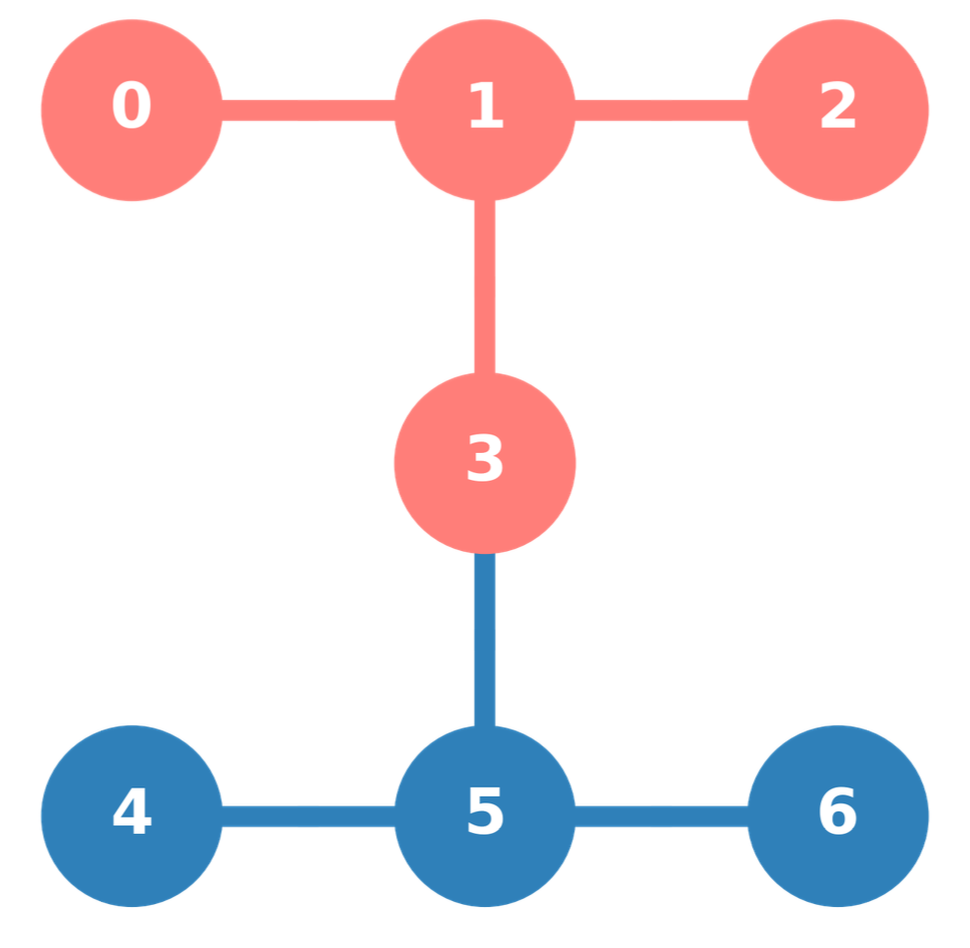}
    \caption{The topology of the NISQ computer used for the experiments. We made use of the seven qubit device ibm perth for both cases M+M=nM and nM+nM=M. The circles represent the qubits, and the lines represent the connectivity between the qubits. For example, qubits three and five are connected, meaning one can perform a CNOT between them directly. The qubits that are coloured in red are used for the experiments, while the blue qubits are not. For the case M+M=nM, we use qubits one to three; for the case nM+nM=M, we use qubits zero to three.}
    \label{Fig21}
\end{figure}

\subsection{Implementation of Quantum Circuits}
For our experiments, the circuits are executed using the NISQ device provided by IBM Quantum Experience (IBMQE). The seven-qubit device called ibm\_perth was used. Fig. \ref{Fig21} gives a diagrammatic representation of the qubits in the quantum computer as well as the connections between the qubits. The connectivity of the qubits plays an essential role in constructing the circuits for the quantum channels. This is because, on all NISQ devices, the quality of the results depends on the number of controlled not gates (CNOTs) used. One should note that this is not the only factor but is most important to us in our simulations. The NISQ devices can only implement a small number of CNOTs before the qubit interactions, noise, and decoherence affect the results. Moreover, the implementation of CNOTs between qubits that are not directly connected requires a SWAP gate to swap the state of the qubits so that the CNOTs can be implemented between directly connected qubits, and then the states are swapped back, which is done automatically by the quantum computer. This leads to problems as SWAP gates are equivalent to three CNOTs, significantly increasing the number of CNOTs. 

We have mitigated this issue by looking at the circuits designed in Section 3 and making some additions whenever we implement controlled operations between qubits that are not directly connected. Rather than letting the quantum computer perform SWAP operations, we manually add the SWAP gate to minimise the number of SWAPs needed, minimising the number of CNOTs used.

For the M+M=nM case, we use qubits one to three in Fig. \ref{Fig21}. Qubit two is the system qubit, and qubits one and three correspond to the environment. For the Markovian channels $\Lambda_{t}^{(1)}$ and $\Lambda_{t}^{(2)}$ there are no controlled operations between unconnected qubits. However, for the non-Markovian total channel $\Lambda_{t}^{(T)}$, we see that we require a controlled operation between qubits three and two. 

\begin{figure}
    \centering
    \includegraphics[scale=0.3]{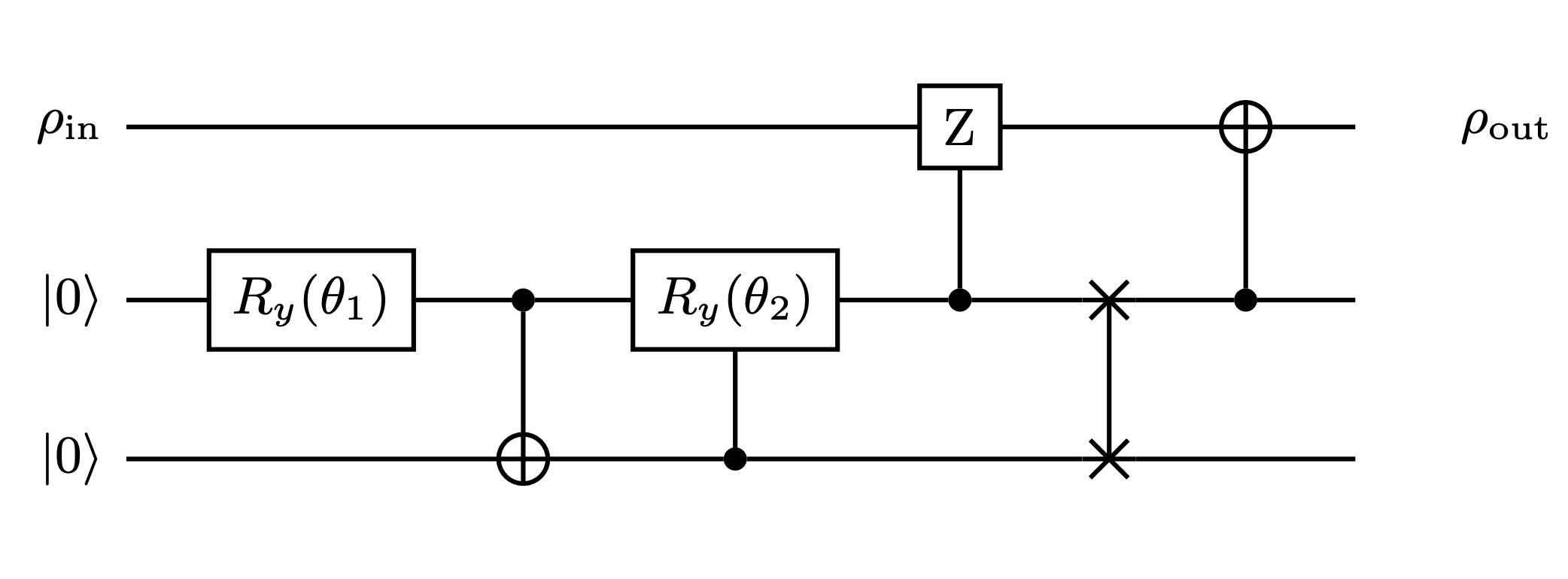}
    \caption{Quantum circuit implementing the total non-Markovian channel $\Lambda_{t}^{(T)}$ for probability $p(t)$ with the addition of a swap gate to mitigate controlled operations between unconnected qubits.}
    \label{fig:circ_nM_CHT}
\end{figure}

Fig. \ref{fig:circ_nM_CHT} shows the new quantum circuit for the non-Markovian channel $\Lambda_{t}^{(T)}$ with the additional SWAP gate.

For the case nM+nM=M, we use qubits one as the system and qubits zero, two and three as the environment. Since there will be no controlled operations between unconnected qubits for the simulation of the channels in the non-Markovian channel addition case, we do not need to modify the circuits from Section 3 with additional SWAP gates.

Once the circuits are executed, a  process tomography is performed on the channels for each time step $\Delta t=0.1 s$ in the interval $t \in [0,3.9]$ for the Markovian channel addition (M+M=nM), and the time interval  $t \in [0,8.9]$ for the non-Markovian channel addition (nM+nM=M). The counts obtained from the results of these circuits are used to generate the process matrices $\chi_{\mathrm{in}}(t)$ for each time step. By adding Gaussian fluctuations, 100 process matrices for each time step were generated. Now, using convex optimisation, as outlined in Section 4, the closest possible process matrices $\chi_{c}(t)$ to the ideal process matrices $\chi_{\mathrm{id}}(t)$ were obtained. Fig. \ref{Fig20} gives an overview of the steps used to construct the quantum circuits, perform the tomography, and reconstruct the channel.

We then pick an intermediate time $s$ such that $t \geq s \geq 0$ and for each pair $(t,s)$, 10000 process matrices are obtained. These are used in the analysis to compute the three metrics. After optimisation, the first metric, the fidelity for the process matrices, measures how close these process matrices are to the ideal. The second metric, the trace norm of the Choi matrix of the channels at each time step,  allows us to check if the simulated channels are trace-preserving. The last metric we considered is the minimum eigenvalue of the Choi matrix of the intermediate map $W(t,s)$, which allows us to characterise the channel as either Markovian or Non-Markovian. These three metrics allow us to verify that the channels simulated are physical and satisfy the case of convexly mixing two Markovian channels to yield a non-Markovian channel and vice versa.

\subsection{Fidelity of the process matrices}

\begin{figure*}[htp]
  \centering
  \includegraphics[scale=0.25]{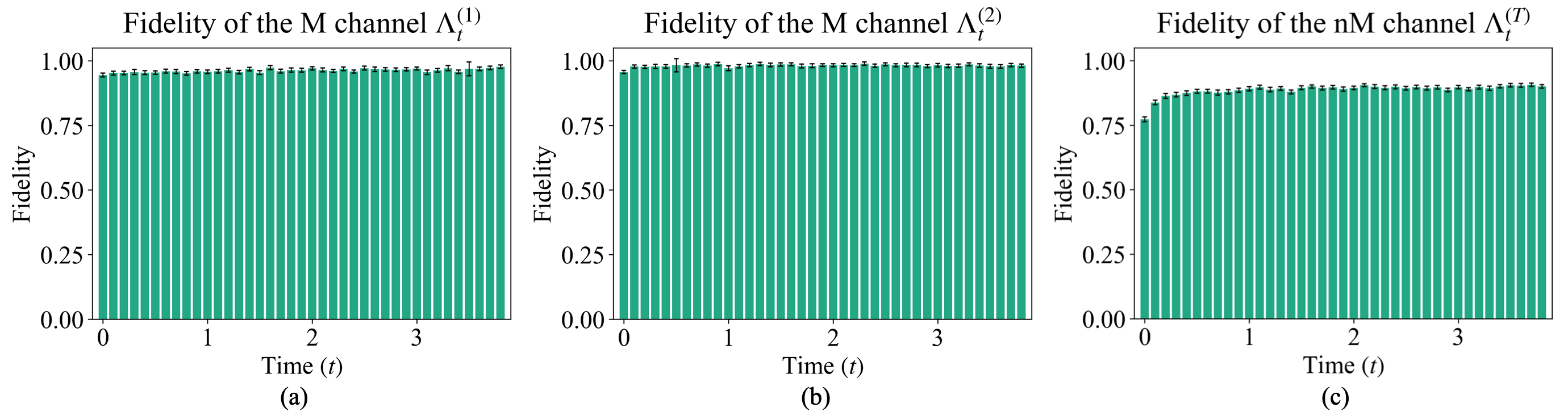}
  \caption{The process fidelities for the $\chi$ matrices of the implemented channels after MLE for times from 0 to 3.9 seconds with time step 0.1 seconds. (a) and (b) show the fidelities for the Markovian channels $\Lambda_{t}^{(1)}$ and $\Lambda_{t}^{(2)}$ respectively. (c) shows the fidelity for the total  non-Markovian channel $\Lambda_{t}^{(T)}$.}
  \label{Fig4}
\end{figure*}

In both cases, we compute the process fidelity of the process matrix for each time $t$ after optimisation, using the following formula  \cite{uriri2020experimental,jozsa1994fidelity}:
\begin{equation}
    \mathrm{F}_{p}(\chi,\chi_{\mathrm{id}})=\frac{\mathrm{Tr}[(\sqrt{\chi}\chi_{\mathrm{id}}\sqrt{\chi})^{1/2}]^{2}}{\mathrm{Tr}[\chi]\mathrm{Tr}[\chi_{\mathrm{id}}]}.
\end{equation}

We note that $\mathrm{F}_{p} \in [0,1]$ when $\mathrm{F}_{p}=1$ this tells us that the process matrix is the same as the ideal, i.e. $\chi=\chi_{\mathrm{id}}$ and when $\mathrm{F}_{p}=0$ the process matrix is far from the ideal process matrix $\chi_{\mathrm{id}}$. We shall compute the process fidelities for the channels $\Lambda_{t}^{(1)}$,$\Lambda_{t}^{(2)}$,$\Lambda_{t}^{(T)}$ for both the cases of Markovian channel addition and non-Markovian channel and plot them for each time step.

In Fig. \ref{Fig4}, we plot the process fidelities for the Markovian channel addition (M+M=nM). We see that in Fig. \ref{Fig4} (a) and (b) the fidelities for the channel $\Lambda_{t}^{(1)}$ and $\Lambda_{t}^{(2)}$ are very close to 1. This tells us that our $\chi$ matrices are very close to the ideal case after MLE. In Fig. \ref{Fig4} (c), we see that the fidelities for the total channel $\Lambda_{t}^{(T)}$ while not as high as the other two channels are still relatively good enough for our experiment. It should be noted that the fidelity is lower in this case due to decoherence and dissipation in the quantum computer. 

\begin{figure*}[htp]
  \centering
  \includegraphics[scale=0.25]{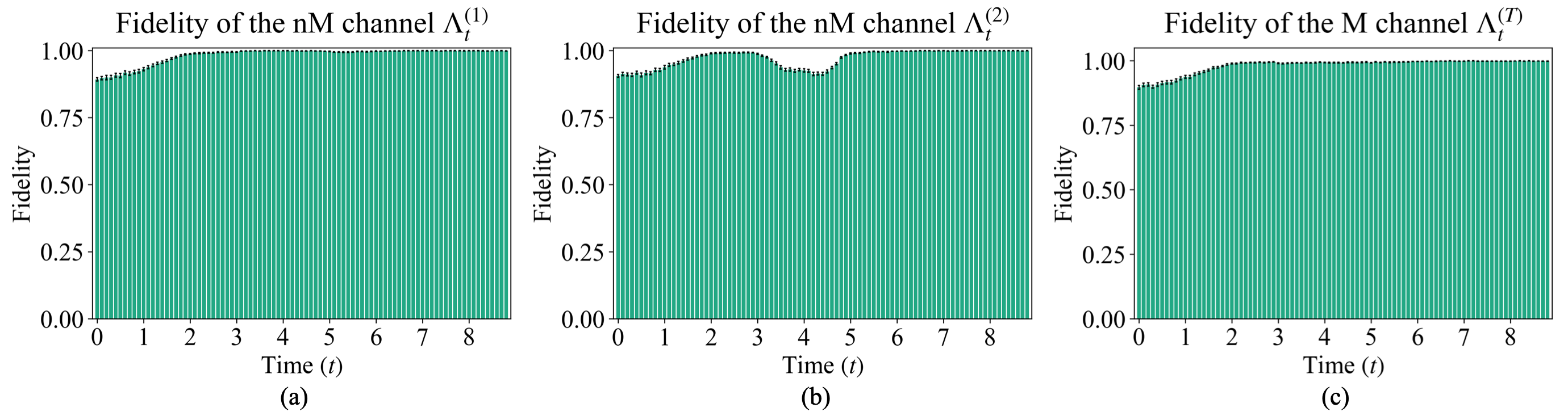}
  \caption{The process fidelities for the $\chi$ matrices of the implemented channels after MLE for times from 0 to 8.9 seconds with time step 0.1 seconds. (a) and (b) show the fidelities for the non-Markovian channels $\Lambda_{t}^{(1)}$ and $\Lambda_{t}^{(2)}$ respectively, the fidelities for these channels are good as for $\Lambda_{t}^{(1)}$ the fidelity is 1 for a large part of the time interval, for $\Lambda_{t}^{(2)}$ the fidelities do fluctuate but are also very good. (c) shows the fidelity for the total Markovian channel $\Lambda_{t}^{(T)}$.}
  \label{Fig6}
\end{figure*}

 We plot the process fidelities for the channels in Fig. \ref{Fig6}. We see that for Fig. \ref{Fig6} (a) and Fig. \ref{Fig6} (b) the fidelities for the individual channels are high and have a value of one for large parts of the time interval, indicating that the quality of the $\chi$ matrices is good. In Fig. \ref{Fig6} (c), the fidelity of the total channel is very good, although at time $t=1.8$ s, the fidelity is low. This tells us that the $\chi$ matrices for the total channel will be close to ideal. For most of the interval but at $t=1.8$ s, the $\chi$ matrix will be far from the ideal matrix. These fidelities are high enough such that the accuracy for the $\chi$ matrices used in our analysis is high.
 
\subsection{Trace Norm of Choi Matrix}
\begin{figure*}[htp!]
  \centering
  \includegraphics[scale=0.25]{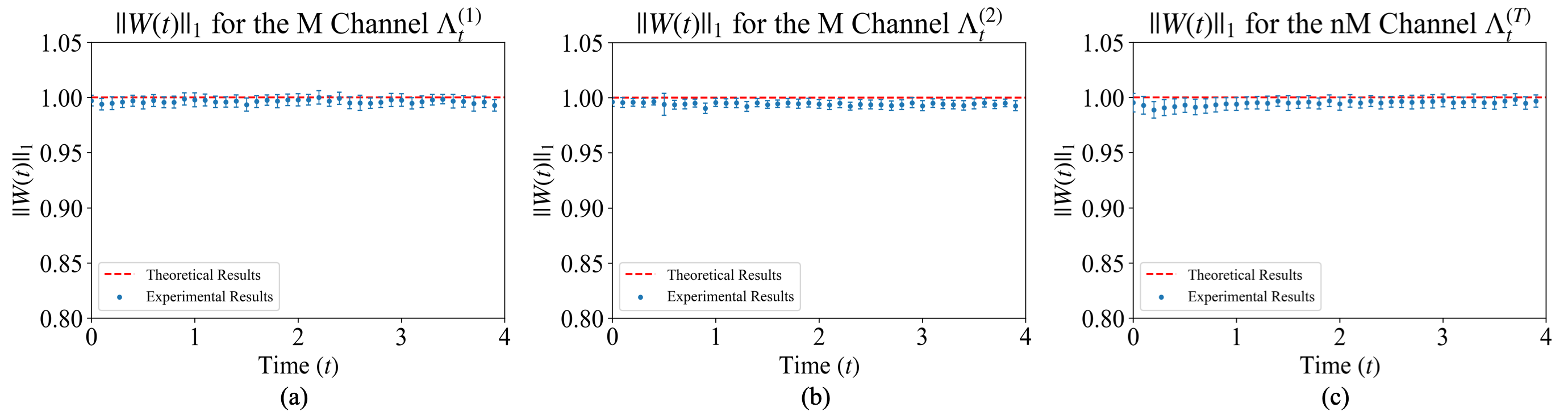}
  \caption{(a)-(c) We plot the trace norm of the Choi matrices of the channels $\Lambda_{t}^{(1)}$, $\Lambda_{t}^{(2)}$ and $\Lambda_{t}^{(T)}$ respectively, for the Markovian channel addition (M+M=nM). For each case, the trace norm is a good enough approximation of 1 to state that the simulated channels are trace-preserving. The deviation and slightly long error bars result from device noise and decoherence.}
  \label{Fig18}
\end{figure*}

The trace norm for the Choi matrix at each time step $t$ can be used to check if the simulated channel was trace-preserving. We calculate the Choi matrix $W(t)$ for each time step by using equations (\ref{transfer_mat_elements})-(\ref{Choi_mat}) from section 4. Once we have found the Choi matrices $W(t)$ we find the trace norm,
\begin{align}
	||W(t)||_{1}=\mathrm{Tr}\left[ \sqrt{W(t)^{\dagger}W(t)}\right].
\end{align}
We know that for any quantum channel $\Lambda_{t}$ to be trace preserving it must satisfy,
\begin{align}
	||W(t)||_{1}=1 \ \forall\ t \geq 0.
\end{align}

\begin{figure*}[htp!]
  \centering
  \includegraphics[scale=0.25]{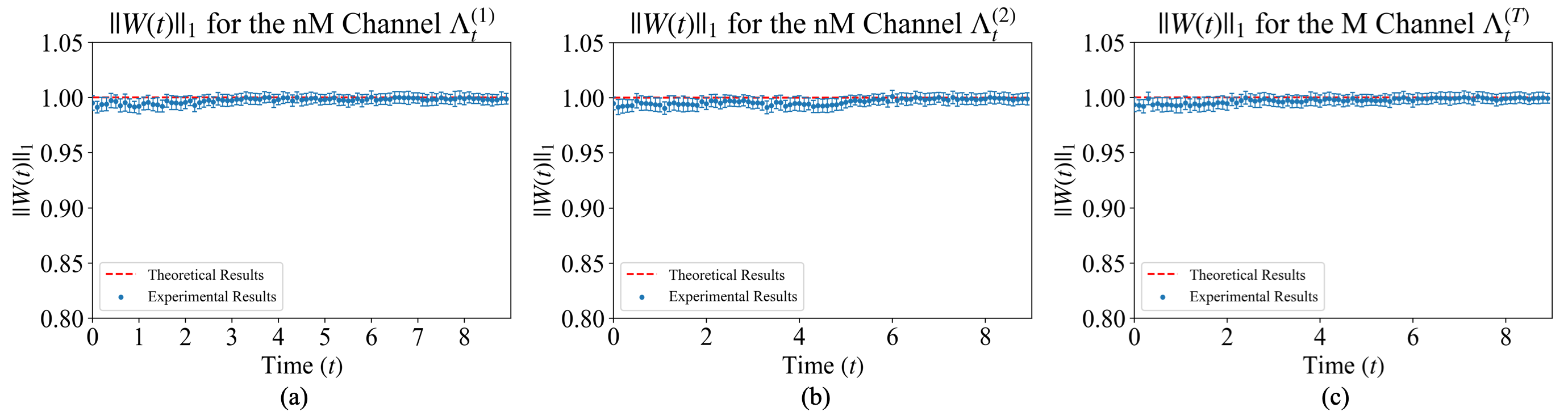}
  \caption{(a)-(c) We plot the trace norm of the Choi matrices of the channels $\Lambda_{t}^{(1)}$, $\Lambda_{t}^{(2)}$ and $\Lambda_{t}^{(T)}$ respectively, for the non-Markovian channel addition (nM+nM=M). The trace norm for these simulations covers the theoretical value and is a good enough approximation to say that the simulated channels were trace-preserving.}
  \label{Fig19}
\end{figure*}

The trace norm for the channels in both cases, M+M=nM and nM+nM=M, are calculated and plotted in Fig. \ref{Fig18} and Fig. \ref{Fig19}, respectively. We see that in Fig. \ref{Fig18} that the trace norm of $W(t)$ for the Markovian channel addition (M+M=nM) is a good enough approximation of 1 for the simulated channel to be considered trace-preserving. It should be noted that device noise and finite sampling are to blame for the deviations from 1. In Fig. \ref{Fig19}, we see that for the case of the non-Markovian channel addition (nM+nM=M), the trace norm of $W(t)$ is a better approximation of 1 as they cover the theoretical curve. The same reasons for the deviations in the Markovian channel addition also cause the deviations seen here. In both cases, the trace norm is a good enough approximation to consider the simulated channels trace-preserving.

\subsection{Minimum Eigenvalue of $W(t,s)$}

\begin{figure*}[htp!]
  \centering
  \includegraphics[scale=0.25]{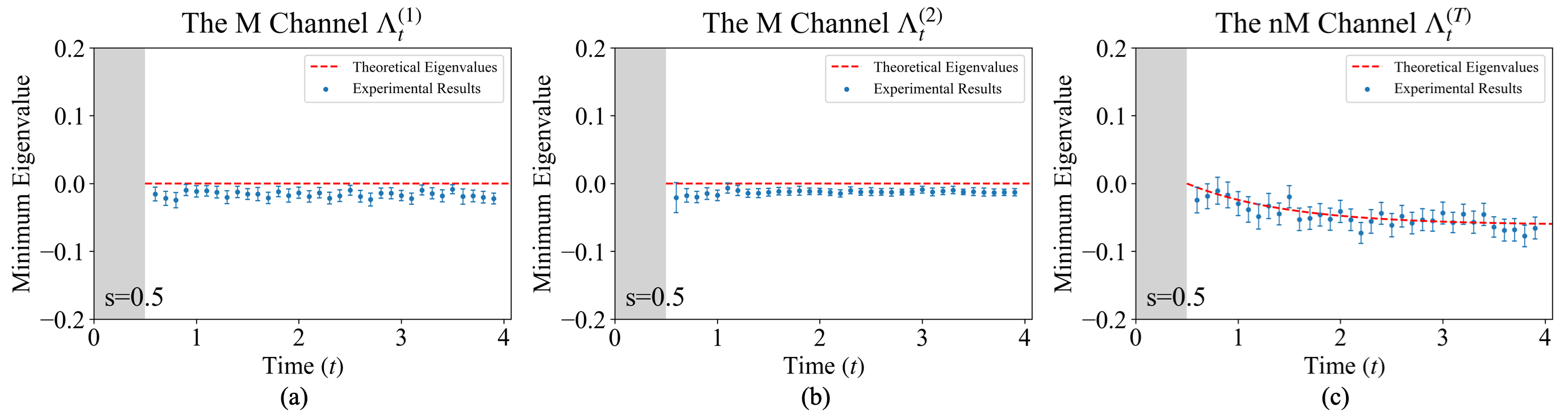}
  \caption{The plot (a) shows the minimum eigenvalues of the IM for the Markovian channel $\Lambda_{t}^{(1)}$, it is clear that the experimental results are a good enough approximation of the zero eigenvalue. (b) Shows the minimum eigenvalues of the IM for the channel $\Lambda_{t}^{(2)}$. The experimental results in this case are better than for the first channel as the standard deviation is small for all the points. (c) shows the minimum eigenvalues of the IM for the total non-Markovian channel $\Lambda_{t}^{(T)}$, the points cover the theoretical curve, hence this channel is definitively non-Markovian by the CP divisibility criteria \cite{rivas2010entanglement}.}
  \label{Fig5}
\end{figure*}

\begin{figure*}[htp!]
  \centering
  \includegraphics[scale=0.25]{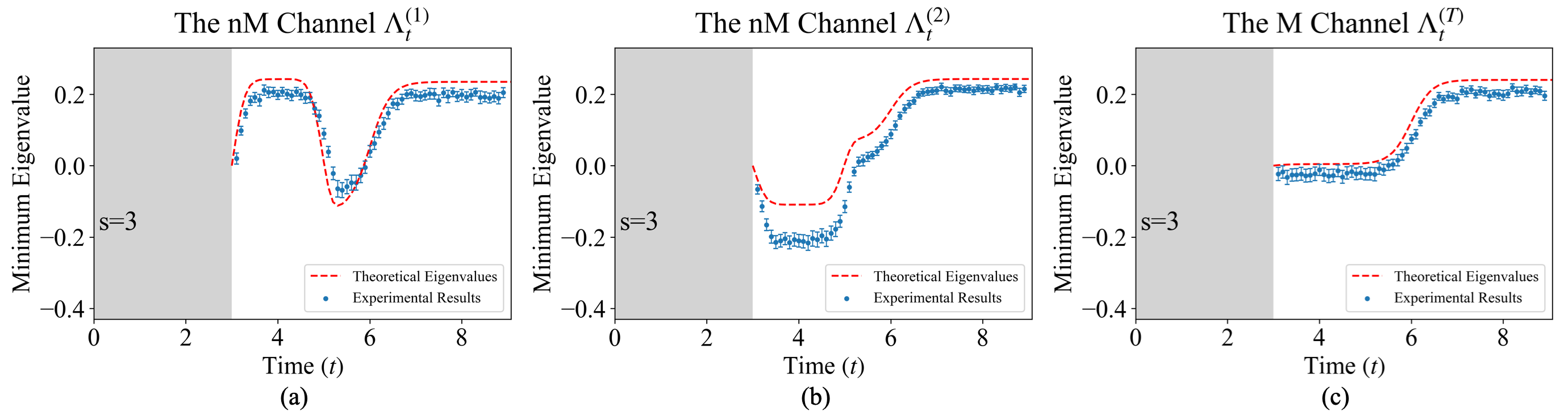}
  \caption{The plot (a) shows the minimum eigenvalues of the IM for the non-Markovian channel $\Lambda_{t}^{(1)}$, it is clear that the experimental results are in agreement with the theoretical minimum eigenvalues of the IM. (b) Shows the minimum eigenvalues of the IM for the channel $\Lambda_{t}^{(2)}$, the experimental results in this case deviate from the theoretical curve but this channel is still non-Markovian by CP divisibility. (c) shows the minimum eigenvalues of the IM for the total Markovian channel $\Lambda_{t}^{(T)}$, the points cover the theoretical curve and the initial points, although negative, are still a good enough approximation of zero, this is a problem face when the minimum eigenvalue is zero. Hence, this channel is Markovian by the CP divisibility criteria \cite{rivas2010entanglement}.}
  \label{Fig7}
\end{figure*}

Now, we use the characterisation method outlined in Section 4 to classify the channels. By plotting the minimum eigenvalues of each Choi Matrix W(t,s) of the IM, we can easily classify the corresponding channel as Markovian if all the minimum eigenvalues are non-negative or Non-Markovian if any minimum eigenvalue is negative. This is according to the CP-divisibility criteria. 

First, we look at the minimum eigenvalues for the Markovian channel $\Lambda_{t}^{(1)}$ $(s= 0.5)$ in Fig. \ref{Fig5} (a). We observe that the minimum eigenvalues are all negative. This differs from the theoretical minimum eigenvalues, which are all zero. This discrepancy is expected since it is difficult to measure zero in any experiment due to standard deviation. This discrepancy can also be seen in Fig. \ref{Fig5} (b) between the theoretical and experimental minimum eigenvalues for the Markovian channel $\Lambda_{t}^{(2)}$ $(s= 0.5)$. In both these cases, the minimum eigenvalues are very small negative values and can be accepted as approximations of zero.  
The minimum eigenvalues for the  Non-Markovian channel $\Lambda_{t}^{(T)}$ $(s= 0.5)$, as shown in Fig. \ref{Fig5} (c), show definitively that this channel is Non-Markovian as the minimum eigenvalues, within their standard deviation, are all negative. While the minimum eigenvalues do fluctuate due to noise, the eigenvalues follow the behaviour of the theoretical curve. 

Next, we look at the minimum eigenvalues for the Non-Markovian channels $\Lambda_{t}^{(1)}$ and 
$\Lambda_{t}^{(2)}$ $(s= 3)$, shown in Fig. \ref{Fig7} (a) and (b). We see that the minimum eigenvalue for $\Lambda_{t}^{(1)}$ is negative in the time interval $t=51$ s to $t=5.8$ s and for $\Lambda_{t}^{(2)}$ the minimum eigenvalue is negative in the interval $t=3$ s to $t=5.4$ s. This shows that these channels are Non-Markovian. We can also see that the minimum eigenvalues follow the theoretical curves. While there are slight deviations from the curve, these deviations can be attributed to noise from the experiment.  
Lastly, in Fig. \ref{Fig7} (c) we look at the minimum eigenvalues for the Markovian channel $\Lambda_{t}^{(T)}$ $(s= 3)$ . We see that the minimum eigenvalues between t=0 and t=6 are negative when the theoretical eigenvalues are zero. Once again, this discrepancy can be attributed to the difficulty of measuring zero in any experiment due to standard deviation. The minimum eigenvalues in this time interval are very small negative values and can be accepted as approximations of zero. Therefore, the channel can be classified as Markovian. 

In each case, we see that the minimum eigenvalues follow the theoretical curves and that the channels can be successfully classified as either Markovian or Non-Markovian using the CP-divisibility criteria.

\section{Conclusion}
We have successfully performed simulations of mixtures of two Markovian single qubit Pauli channels that give rise to a non-Markovian single qubit Pauli channel and two non-Markovian single qubit depolarising channels that give rise to a Markovian single qubit depolarising channel. The success of these simulations has been verified in three ways. Firstly, we have shown by the CP-divisibility criteria that the channel reconstructed from the simulations of the M+M=nM mixtures and the nM+nM=M mixtures are non-Markovian and Markovian, respectively. We have also shown that the reconstructed channels have high fidelity to the theoretical channel. Lastly, we have shown that the Choi matrices have a required trace norm of one, indicating that the simulated channels are physical. This demonstrates the effectiveness of designing circuits with NISQ device topology in mind and provides more accurate results, as the quantities we use to benchmark our simulation are all excellent enough approximations of their theoretical values. This demonstrates that using the least squared objective function \cite{huang2020reconstruction} and convex optimization is a better choice for reconstructing our channel than MLE. The success of these experiments shows that a NISQ computer can be used to simulate more complex open quantum systems. However, future work is needed to verify that the strategies used here generalise to more general quantum channels.
In this work, we consider only single qubit Pauli channels. The simulation of many qubit and non-Pauli channels is not considered here but will be examined in future work. We would like to see if our effective strategies in simulating single qubit channels can aid in simulating single qubit non-Pauli channels and many qubit Pauli and non-Pauli channels. 
Future work could look at applying the developed experimental pipeline to different channels since the pipeline is not specific to the channels used in this work. Other future work includes testing and comparing different objective functions for optimization and changing the constraints. One could also look into using a semi-definite program for the optimization part of the reconstruction.


\bmhead{Acknowledgments}
 We would like to thank Ms. S. M. Pillay for her helpful discussions and assistance in proofreading the manuscript.
\section*{Declarations}

\begin{itemize}
\item Funding:
This work is based upon research supported by the
National Research Foundation of the Republic of South
Africa. Support from the NICIS (National Integrated
Cyber Infrastructure System) e-research grant QICSA is
kindly acknowledged.

\item Conflict of interest/Competing interests: 
Francesco Petruccione the Chair of the Scientific Board and Co-Founder of
QUNOVA computing. The authors declare no other competing interests.

\item Ethics approval:
Not applicable.
\item Consent to participate:
Not applicable.
\item Consent for publication:
Not applicable.
\item Availability of data and materials:
Not applicable.
\item Code availability:
The codes used to generate all the data for this research are available from the corresponding author upon reasonable request.
\item Authors' contributions:
IJD and IS came up with the idea for the project. IJD implemented the code and wrote up the manuscript. All Authors discussed the results and reviewed the manuscript.

\end{itemize}

\begin{appendices}

\section{Intuition for the choice of functions in the non-Markovian channel addition}\label{secA1}

Choosing the functions $q(t)$ and $r(t)$ for the non-Markovian channel addition was a non-trivial task. We shall provide some intuition on how these functions were chosen and the logic behind these choices. The calculations in section 2 give us the following conditions on the functions $q(t)$ and $r(t)$:
\begin{align}
   & 0\leq q(t) < 1 \hspace{2mm} \mathrm{and} \hspace{2mm} q(0)=0,\nonumber\\
   \nonumber\\
   &0\leq r(t) <1\hspace{2mm} \mathrm{and} \hspace{2mm} r(0)=0.
\end{align}
Now from section 2 we have that for the channels $\Lambda_{t}^{(1)}$ and $\Lambda_{t}^{(2)}$ to be non-Markovian their respective decay rates should be negative for some time interval. From equation (\ref{decay_rates_non_markov_channels}), we see that:
\begin{align}
    &\dot{q}(t') <0 \hspace{2mm} \mathrm{for} \hspace{1mm}\mathrm{some}\hspace{2mm} t' \geq 0\nonumber\\
    \nonumber\\
    &\dot{r}(t'') <0 \hspace{2mm} \mathrm{for} \hspace{1mm}\mathrm{some}\hspace{2mm} t'' \geq 0.
\end{align}
Equation (\ref{decay_rate_total_Markov}) tells us that for the total channel $\Lambda_{t}^{(T)}$ to be Markovian, we must have:
\begin{equation}
    h\dot{q}(t)+(1-h)\dot{r}(t) \geq 0 \hspace{2mm} \forall t\geq 0,
\end{equation}
where $h \in [0,1]$. Now setting $h=\frac{1}{2}$ as in section 3, we get the decay rate for the total channel as $\frac{1}{2}(\dot{q}(t)+\dot{r}(t))$. The intuition behind how to choose $q,r$ is as follows. We need to choose the functions $q(t),r(t)$ such that when $\dot{q}(t') < 0$ on some interval $t' \in [a,b]$, then $\dot{r}(t') >0$ for $t' \in [a,b]$ and vice versa. This ensures that the convex sum $\frac{1}{2}(\dot{q}(t)+\dot{r}(t)) \geq 0$ for all times $t$ and will satisfy the conditions of non-Markovianity for the individual channels, i.e. equation (A2). We shall parameterize the functions $q(t)$ and $r(t)$ as follows:
\begin{align}
    q(t)&=a(t)+b(t)\nonumber\\
    r(t)&=a(t)-b(t),
\end{align}
and from equation (A1) we see that $a(0)=b(0)=0$ as well as,
\begin{align}
    b(t) \leq a(t) < 1-b(t).
\end{align}
Taking the derivatives of $q$ and $r$ we get,
\begin{align}
    \dot{q}(t)&=\dot{a}(t)+\dot{b}(t)\nonumber\\
    \dot{r}(t)&=\dot{a}(t)-\dot{b}(t),
\end{align}
and taking their convex mixture yields,
\begin{equation}
    \frac{1}{2}(\dot{q}(t)+\dot{r}(t))=\dot{a}(t).
\end{equation}
From equations (A5)-(A7) we have the conditions on the functions $a(t)$ and $b(t)$. We note that since $a(t)$ is bounded between the functions $b(t)$ and $1-b(t)$, if we choose $b(t)$ to have the shape of a Plank distribution where $b(0)\approx0$ then $a(t)$ needs to satisfy $a(0)\approx0$ and equation (A5). We choose $a(t)$ as the sum of two sigmoid functions to satisfy the conditions on $a(t)$. Now that we have the general shape of both $a(t)$ and $b(t)$ by using translation and scaling factors, we can transform the general shapes of these functions to satisfy all the bounds in equations (A5)-(A7). Hence the choice of the functions $a(t)$ and $b(t)$ in equation (\ref{functions_a_b}). This is the intuition behind the design of the non-Markovian channels in the (nM+nM=M) experiment. A similar approach can be followed for designing other experiments.
\end{appendices}


\bibliography{References}

\end{document}